\UseRawInputEncoding
\documentclass[%
 reprint,
 amsmath,amssymb,
 aps,prb, twocolumn, showpacs
]{revtex4-1}

\usepackage{graphicx}
\usepackage{dcolumn}
\usepackage{bm}
\usepackage{float}
\usepackage{caption}
\usepackage{subfigure}
\usepackage{amsmath}
\usepackage{mathrsfs}
\usepackage{amssymb}
\usepackage{hyperref}
\usepackage{xcolor}

\begin{document}

\preprint{}

\title{Ergodicity in glass relaxation}

\author{Li Wan}
\email{lwan@wzu.edu.cn}
\affiliation{Department of Physics, Wenzhou University, Wenzhou 325035, P. R. China}
\date{\today}

\begin{abstract}
We derive an equation for the glass relaxation. In the derivation, the Zwanzig-Mori projection method is not applied explicitly, which makes our equation different from the mode coupling theory. Due to the nonlinearity, it is difficult to solve the equation to get the full behaviors of the glass relaxation. But we can simplify the equation when time approaches infinity and obtain the static result analytically. The static result shows that the density correlation function decays to zero finally, meaning that the glass relaxation is ergodic. In this study, we also find that the force fluctuation of one individual particle averaged in the glass is sensitive to the temperature and is suggested to be a parameter to reflect the structural transition for the glass relaxation. 
\begin{description}
\item[Keywords]
{glass relaxation; mode coupling theory; force fluctuation; density correlation function; ergodicity}
\end{description}
\end{abstract}

\maketitle

\section{Introduction}
\label{I}
In the cooling of a liquid below its melting temperature, the liquid normally falls into its stable thermodynamic state to form a crystal. However, when the cooling rate is rapid enough to avoid the crystallization, the liquid can remain structurally disordered below its melting temperature known as the supercooled liquid(SL)~\cite{Kob,Pablo,Gotz,Donth}. Particles in the SL move with time and change the initial configuration of the SL to a new one. The overlap of the new configuration with the initial configuration of the SL is averaged in ensemble and is used to determine the degree of the relaxation. In the SL, the overlap decays to zero with the time evolving, meaning that the SL experiences all the possible configurations. The relaxation of the SL then is ergodic. Generally, the time for the decaying of the overlap to zero is defined as the relaxation time. It is believed that further cooling makes the SL experience a phase transition and change to be a glass below a phase transition temperature~\cite{Kob,Pablo,Gotz,Donth,Zallen}. In the glass, the physical properties are expected to be different from those of the SL. However, it is still not clear what the glass is and how to identify the phase transition between the SL and the glass~\cite{Pablo,Angell}. Investigating these problems has given rise to thick literatures with many theoretical perspectives and approaches suggested~\cite{Berthier,Sciortino,Dyre,Chen,Adam,Parisi,Lubchenko,Cavagna,Leutheusser,Kauzmann,Chandler,Tarjus,Ediger,Langer}. One of the topics in the investigation is to understand whether the overlap of the configurations in the glass decays to zero or not. If not, the relaxation time is divergent, and the relaxation is non-ergodic. It is known that the relaxation time increases with the decreasing of cooling temperature. The relaxation time of the glass is very long if exists, and exceeds the time scale accessed by experiments or simulations. Thus, it is impossible for experiments or simulations to judge if the glass relaxation is ergodic or non-ergodic. But, it is possible to be touched by theories. In this paper, we propose a theory for this goal.\\

One of the powerful theories to investigate the relaxation of a SL is the mode coupling theory(MCT)~\cite{Gotz,Leutheusser}. The MCT applies the Zwanzig-Mori projection method to project physical variables of the SL onto its slow variables~\cite{Zwanzig,Reichman,Janssen}. In this way, a General Langevin Equation(GLE) can be derived as the basis of the MCT. In the GLE, fast variables of the SL have been dropped off and slow variables remain to grasp the relaxation of the SL in the long time evolving. By the projection method, a memory kernel is obtained in the GLE, which is originated from the components of fast variables perpendicular to the slow variables. The memory kernel shows that the history of the relaxation influences the future relaxation of the system. The derivation for the GLE is rigorous, but it is impossible to solve the GLE due to the memory kernel. The MCT provides some approximations to simplify the GLE for numerical calculations. Especially, the memory kernel in the GLE has an factor $e^{i(1-\hat{P})\hat{L}t}$ with $\hat{P}$ the projection operator, $\hat{L}$ the Liouville operator, $i$ the imaginary unit and $t$ the time. This factor in the memory kernel is not accessible and has to be replaced with $\hat{P}'e^{i\hat{L}t}\hat{P}'$ directly in the MCT for the numerical calculation with the operator $\hat{P}'$ constructed by slower variables~\cite{Reichman}. Such replacement of the operator is not under control and without any approximation to be shown.\\

In the MCT, density correlations have been adopted as slow variables to reflect the overlap of the configurations in the SL. The numerical results of the MCT show the cage effect for the SL relaxation successfully~\cite{Gotz}. The cage effect means that particles are enclosed by cages in the SL and difficult to escape from the cages. The degree of the difficulty for the particles escaping from the cages is enhanced when the cooling temperature is decreased, and the relaxation time is lengthened as well. Until it reaches a temperature noted as the MCT temperature, the density correlation decays to a constant of nonzero and the relaxation time is divergent. That means below the MCT temperature, the relaxation of the SL is changed to be non-ergodic. We emphasize here that the MCT temperature is still higher than the temperature of the SL-glass phase transition defined by Angell plot~\cite{Angell1,Janssen}. Thus, the MCT is not applicable in the temperature range of the glass. And the non-ergodic phenomena revealed by the MCT have never been observed by the experiments~\cite{Fuchs,Berthier}. In order to extend the MCT to the temperatures below the MCT temperature, many mechanisms have been introduced to modify the GLE, such as projection on the slower variables, or hopping of the particles from the cages, etc~\cite{Das,Kim,Nishino,Bengtzelius,Szamel,Janssen1,Biroli,Adam,Charbonneau}. These modifications make the MCT much more complicated and still can not answer the question if the relaxation is ergodic or non-ergodic, because the time for the numerical calculations is limited. What's more, the replacement of the operator in the MCT as we have mentioned before has never been justified in the modifications.\\

In this study, we derive an equation different to the GLE in the MCT. In our derivation, the Zwanzig-Mori projection method is not applied explicitly to avoid the replacement of the operator. Our equation can be simplified when time approaches infinity. In this way, the static result is obtained analytically. By using the static result, we can answer the question if the relaxation of the glass is ergodic or non-ergodic. \\

\section{theory}
We consider a general system comprised of particles of various species. The particle number of the $l$-th species is denoted by $N_l$. The total particle number is denoted by $N$. In this study, only the pairwise interactions between the particles are considered.\\

\subsection{density-density correlation function}
\label{IIA}
We start our study from Newton's Equations of Motion
\begin{align}
\label{IIA1}
&\frac{d \vec{x}_n^{(l)}}{dt}=\vec{v}_n^{(l)},\nonumber\\
&M^{(l)}\frac{d \vec{v}_n^{(l)}}{dt}=-\sum_{p,m}\frac{\partial V(\vec{x}_n^{(l)},\vec{x}_m^{(p)})}{\partial \vec{x}_n^{(l)}}.
\end{align}
Here, $\vec{x}_n^{(l)}$ is the displacement of the $n$-th particle of the $l$-th species with the index number $n$ from $1$ to $N_l$. $\vec{v}_n^{(l)}$ is the velocity for the particle of $\vec{x}_n^{(l)}$ with the mass of $M^{(l)}$. The potential $V(\vec{x}_n^{(l)},\vec{x}_m^{(p)})$ is for the pairwise interaction between the two particles of $\vec{x}_n^{(l)}$ and $\vec{x}_m^{(p)}$.\\

The number density of particles at the displacement of $\vec{x}$ for the $l$-th species is introduced as $\rho^{(l)}(\vec{x})=\sum_n \delta(\vec{x}-\vec{x}^{(l)}_n)$ and the velocity density at $\vec{x}$ for the $l$-th species is $\vec{J}^{(l)}(\vec{x})=\sum_n \vec{v}^{(l)}_n\delta(\vec{x}-\vec{x}^{(l)}_n)$. Here, $\delta$ is the Dirac's delta function. We apply the Fourier transformation on $\rho^{(l)}(\vec{x})$ and $\vec{J}^{(l)}(\vec{x})$ to getting quantities $\rho^{(l)}_{\vec{k}}$ and $\vec{J}_{\vec{k}}^{(l)}$ respectively in the reciprocal space. The subscript $\vec{k}$ of the quantities is the wave vector in the reciprocal space. We refer to $\rho^{(l)}_{\vec{k}}$ as density and $\vec{J}^{(l)}_{\vec{k}}$ as current. For convenience, we introduce a factor of $1/\sqrt{N_l}$ and rewrite the quantities as
\begin{align}
\label{IIA2}
&\rho^{(l)}_{\vec{k}}=\frac{1}{\sqrt{N_l}}\sum_n e^{i\vec{k}\cdot \vec{x}_n^{(l)}},\nonumber\\
&\vec{J}_{\vec{k}}^{(l)}=\frac{1}{\sqrt{N_l}}\sum_n \vec{v}_n^{(l)}e^{i\vec{k}\cdot \vec{x}_n^{(l)}}.
\end{align} 
We make time derivatives of the two quantities $\rho^{(l)}_{\vec{k}}$ and $\vec{J}_{\vec{k}}^{(l)}$ to get the equations for the $l$-th species. Before we show the equations, we non-dimensionalize the quantities for convenience. We assign one arbitrarily chosen species of the system as the first species, and take the physical quantities of the first species as the references. Quantities of all the other species are normalized on the references. Thus, the normalized mass for a particle of the $l$-th species is $\alpha^{(l)}=M^{(l)}/M^{(1)}$. Since we will use the Lennard-Jones(LJ) potential for the Molecular Dynamics(MD) simulation later, here we take the LJ potential as an example for the non-dimensionalization. The LJ potential takes the form of 
\begin{align}
\label{IIA3}
V(\vec{x}_n^{(l)},\vec{x}_m^{(p)})=4\epsilon^{(l,p)}\left[\left(\frac{\sigma^{(l,p)}}{r}\right)^{12}-\left(\frac{\sigma^{(l,p)}}{r}\right)^{6}\right]
\end{align}
with $r=|\vec{x}_n^{(l)}-\vec{x}_m^{(p)}|$ the distance between the two particles of $\vec{x}_n^{(l)}$ and $\vec{x}_m^{(p)}$. Here, $\epsilon^{(l,p)}$ and $\sigma^{(l,p)}$ are the LJ parameters for the interaction between the $l$-th and the $p$-th species. The notations of $\epsilon^{(l,l)}$ and $\sigma^{(l,l)}$ are simplified to be $\epsilon^{(l)}$and $\sigma^{(l)}$ respectively. We take $\epsilon^{(1)}$ as the energy scale and $\sigma^{(1)}$ as the length scale. Then we introduce a velocity scale $v_{scale}$ by $v_{scale}^2=\epsilon^{(1)}/M^{(1)}$ and a time scale by $t_{scale}=\sigma^{(1)}/v_{scale}$. We normalize the velocity $\vec{v}$ of particles by $v_{scale}$, the time $t$ by $t_{scale}$ and the wave vectors $\vec{k}$ by $1/\sigma^{(1)}$. Finally, we introduce an normalized temperature $\eta=k_BT_{SI}/\epsilon^{(1)}$ with $T_{SI}$ the temperature of the system in the SI unit of Kelvin, and $k_B$ the Boltzmann factor. After the non-dimensionalization, we still keep the notations of $\vec{k}$, $t$, $\vec{x}$, $\vec{v}$ and $V$ to save symbols. We state that all the equations in the following have been non-dimensionalized.\\

By using Eq.(\ref{IIA1}), we make time derivatives of $\rho^{(l)}_{\vec{k}}$ and $\vec{J}_{\vec{k}}^{(l)}$ to get the equations
\begin{align}
\label{IIA4}
&\frac{d \rho_{\vec{k}}^{(l)}}{dt}=i\vec{k}\cdot \vec{J}_{\vec{k}}^{(l)},\nonumber \\
&\frac{d \vec{J}_{\vec{k}}^{(l)}}{dt}=\sum_{p}F_{\vec{k}}^{(l,p)}+\sum_n \vec{v}_n^{(l)}e^{i\vec{k}\cdot \vec{x}_n^{(l)}}(i\vec{k}\cdot \vec{v}_n^{(l)})/\sqrt{N_l}
\end{align}
with $F_{\vec{k}}^{(l,p)}=-\sum_{n,m} e^{i\vec{k}\cdot \vec{x}_n^{(l)}}[\frac{\partial V(\vec{x}_n^{(l)},\vec{x}_m^{(p)})}{\partial \vec{x}_n^{(l)}}]/[\alpha^{(l)}\sqrt{N_l}]$ originated from the force. We combine the above two equations by canceling $\vec{J}_{\vec{k}}^{(l)}$, and then have a second order differential equation, which reads 
\begin{align}
\label{IIA5}
\frac{d^2 \rho_{\vec{k}}^{(l)}}{dt^2}=&\sum_{p}\left\{i\vec{k}\cdot F_{\vec{k}}^{(l,p)}\right\}\nonumber\\
&+\sum_n \left\{(i\vec{k}\cdot \vec{v}_n^{(l)})e^{i\vec{k}\cdot \vec{x}_n^{(l)}}(i\vec{k}\cdot \vec{v}_n^{(l)})/\sqrt{N_l}\right\}.
\end{align}
For clarity, we note $R_{\vec{k}}^{(l)}$ as the value of $\rho_{\vec{k}}^{(l)}$ at the initial time and note $T_{\vec{k}}^{(l,p)}$ as the initial value of $F_{\vec{k}}^{(l,p)}$. Then we define a density-density correlation function $<\rho^{(l)}_{\vec{k}}R^{(w)}_{-\vec{k}}>$ in which the time dependent density $\rho^{(l)}_{\vec{k}}$ is correlated to the initial density $R^{(w)}_{-\vec{k}}$. For short, we refer to $<\rho^{(l)}_{\vec{k}}R^{(w)}_{-\vec{k}}>$ as the density correlation function(DCF). In the DCF, the wave vector $\vec{k}$ of $\rho_{\vec{k}}^{(l)}$ is opposite to the wave vector $-\vec{k}$ of $R^{(w)}_{-\vec{k}}$ in direction. In this way, the momentum conservation of $<\rho^{(l)}_{\vec{k}}R^{(w)}_{-\vec{k}}>$ is guaranteed due to the translational invariance of the system.\\

We correlate the initial density $R^{(w)}_{-\vec{k}}$ on both sides of Eq.(\ref{IIA5}) to get a differential equation for $<\rho^{(l)}_{\vec{k}}R^{(w)}_{-\vec{k}}>$. On the right hand side of the equation, we have two terms. The second term reads $\sum_n <(i\vec{k}\cdot \vec{v}_n^{(l)})e^{i\vec{k}\cdot \vec{x}_n^{(l)}}(i\vec{k}\cdot \vec{v}_n^{(l)})R_{-\vec{k}}^{(w)}>/\sqrt{N_l}$, which can be simplified by using the Theorem of Equipartition Energy(TEE) $<(\vec{k}\cdot \vec{v}_n^{(l)})^2>\approxeq \eta k^2/\alpha^{(l)}$ and with some approximations implemented. Here, $k$ is the magnitude of $\vec{k}$. After the simplification, the differential equation is 
\begin{align}
\label{IIA6}
\frac{d^2 <\rho_{\vec{k}}^{(l)}R_{-\vec{k}}^{(w)}>}{dt^2}=&\sum_{p}<(i\vec{k}\cdot F_{\vec{k}}^{(l,p)})R_{-\vec{k}}^{(w)}>\nonumber\\
&-\frac{\eta k^2}{ \alpha^{(l)}}<\rho^{(l)}_{\vec{k}}R^{(w)}_{-\vec{k}}>.
\end{align}
The normalized temperature $\eta$ in Eq.(\ref{IIA6}) is obtained from the TEE. The details for the simplification on the second term can be found in Appendix A. \\

\subsection{force-density correlation function}
\label{IIB}
The first term on the right hand side of Eq.(\ref{IIA6}) is for the correlation between force and density. We will find an equation for this term in this subsection. The property of the Liouville operator reveals
\begin{align}
\label{IIB7}
\frac{d <(\vec{k}\cdot F_{\vec{k}}^{(l,p)})R_{-\vec{k}}^{(w)}>}{dt}=-<(\vec{k}\cdot F_{\vec{k}}^{(l,p)})\dot{R}_{-\vec{k}}^{(w)}>
\end{align}
in page 270 of Ref.(\onlinecite{Hansen}). Here, $\dot{R}_{-\vec{k}}^{(w)}$ means the value of $d \rho_{-\vec{k}}^{(w)}/dt$ at the initial time. We apply the property of the Liouville operator twice, getting
\begin{align}
\label{IIB8}
\frac{d^2 <(\vec{k}\cdot F_{\vec{k}}^{(l,p)})R_{-\vec{k}}^{(w)}>}{dt^2}=<(\vec{k}\cdot F_{\vec{k}}^{(l,p)})\ddot{R}_{-\vec{k}}^{(w)}>.
\end{align}
In the above equation, the factor $\ddot{R}_{-\vec{k}}^{(w)}$ equals the value of $d^2 \rho_{-\vec{k}}^{(w)}/dt^2$ at the initial time, which can be obtained by Eq.(\ref{IIA5}) at the initial time. For $\ddot{R}_{-\vec{k}}^{(w)}$, all the quantities in the right hand side of Eq.(\ref{IIA5}) take the initial values with $\vec{k}$ replaced by $-\vec{k}$, $F_{-\vec{k}}^{(l,p)}$ replaced by $T_{-\vec{k}}^{(l,p)}$, and $l$ replaced by $w$. We utilize Eq.(\ref{IIA5}) to get the expression of $\ddot{R}_{-\vec{k}}^{(w)}$, and substitute the expression in the right hand side of Eq.(\ref{IIB8}). Then we have
\begin{align}
\label{IIB9}
\frac{d^2 <(\vec{k}\cdot F_{\vec{k}}^{(l,p)})R_{-\vec{k}}^{(w)}>}{dt^2}&=\sum_{q}<(\vec{k}\cdot F_{\vec{k}}^{(l,p)})(-i\vec{k}\cdot T_{-\vec{k}}^{(w,q)})>\nonumber\\
&-\frac{\eta k^2}{ \alpha^{(w)}}<(\vec{k}\cdot F_{\vec{k}}^{(l,p)})R^{(w)}_{-\vec{k}}>.
\end{align}
The second term on the right hand side of Eq.(\ref{IIB9}) has been simplified with the application of the TEE and some approximations implemented. The simplification is the same as we have done in Appendix A. 

\subsection{differential equation}
\label{IIC}
Combining Eq.(\ref{IIA6}) and Eq.(\ref{IIB9}) and dropping off the terms related to $<(\vec{k}\cdot F_{\vec{k}}^{(l,p)})R_{-\vec{k}}^{(w)}>$, we derive a fourth order differential equation for $<\rho_{\vec{k}}^{(l)}R_{-\vec{k}}^{(w)}>$. Before we introduce the equation, we define $\mathcal{G}_{\vec{k}}^{(l,w)}=<\rho_{\vec{k}}^{(l)}R_{-\vec{k}}^{(w)}>/<R_{\vec{k}}^{(l)}R_{-\vec{k}}^{(w)}>$ for normalization and write $\vec{k}=k\hat{k}$ with $\hat{k}$ the unit vector along the direction of $\vec{k}$. Then, the equation reads
\begin{align}
\label{IIC10}
\frac{d^4 \mathcal{G}_{\vec{k}}^{(l,w)}}{dt^4}+\zeta_1\cdot \frac{d^2 \mathcal{G}_{\vec{k}}^{(l,w)}}{dt^2}+\zeta_2\cdot \mathcal{G}_{\vec{k}}^{(l,w)}=I_1\cdot I_2
\end{align}
with
\begin{align}
&\zeta_1=\frac{1}{\alpha^{(l)}}+\frac{1}{\alpha^{(w)}},~~~~~~~~~~~~\zeta_2=\frac{1}{ \alpha^{(l)}\alpha^{(w)}},\nonumber\\
&I_1=\frac{1}{\eta^2k^2}\frac{<\mathcal{T}_{\vec{k}}^{(l)}\mathcal{T}_{-\vec{k}}^{(w)}>}{<R_{\vec{k}}^{(l)}R_{-\vec{k}}^{(w)}>},~~~~I_2=\frac{<\mathcal{F}_{\vec{k}}^{(l)}\mathcal{T}_{-\vec{k}}^{(w)}>}{<\mathcal{T}_{\vec{k}}^{(l)}\mathcal{T}_{-\vec{k}}^{(w)}>}\nonumber.
\end{align}
The coefficients $\zeta_1$ and $\zeta_2$ are functional of the masses $\alpha^{(l)}$ and $\alpha^{(w)}$. In $I_1$ and $I_2$, the notations are $\mathcal{T}_{-\vec{k}}^{(w)}=\sum_{q}(-i\hat{k}\cdot T_{-\vec{k}}^{(w,q)})$ and 
$\mathcal{F}_{\vec{k}}^{(l)}=\sum_{p}(i\hat{k}\cdot F_{\vec{k}}^{(l,p)})$. In the above derivation, a factor of $\sqrt{\eta}k$ has been absorbed in $t$ to show the intrinsic role of the temperature $\eta$ for the glass relaxation. It is shown that the temperature $\eta$ appears explicitly only in the right hand side of Eq.(\ref{IIC10}), which is split into two factors  $I_1$ and $I_2$. In $I_1$, the denominator $<R_{\vec{k}}^{(l)}R_{-\vec{k}}^{(w)}>$ is originated from the definition of $\mathcal{G}_{\vec{k}}^{(l,w)}$. We introduce $<\mathcal{T}_{\vec{k}}^{(l)}\mathcal{T}_{-\vec{k}}^{(w)}>$ as the numerator in $I_1$ and the denominator in $I_2$. The introduction of $<\mathcal{T}_{\vec{k}}^{(l)}\mathcal{T}_{-\vec{k}}^{(w)}>$ in Eq.(\ref{IIC10}) will show later that the temperature $\eta$ does not appear explicitly in $I_2$. Thus, the role of the temperature in the glass relaxation is reflected only by the factor $I_1$. The factors $<\mathcal{T}_{\vec{k}}^{(l)}\mathcal{T}_{-\vec{k}}^{(w)}>$ and $<R_{\vec{k}}^{(l)}R_{-\vec{k}}^{(w)}>$ in $I_1$ are obtained from the ensemble average of the system in equilibrium, which can be accessed by MD simulations.\\

The factor $I_2$ is for the force-force correlation, playing the similar role as the memory kernel in the GLE. As we have mentioned in Sec.(\ref{I}), the memory kernel in the GLE can not be calculated. The MCT replaces the memory kernel in the GLE with some other operators for numerical calculation. The replacement is not under control. Comparably, in Eq.(\ref{IIC10}), the correlation $I_2$ can be treated with details shown clearly. The numerical calculation to Eq.(\ref{IIC10}) then is feasible. To demonstrate our theory in a simple way, the physical properties of the glass are assumed to be homogeneous. Then the physical quantities in the reciprocal space depend on only the magnitude $k$ instead of the direction $\hat{k}$  of the wave vector $\vec{k}$.\\

\subsection{factor $I_1$}
\label{IID}
The DCF $<R_{\vec{k}}^{(l)}R_{-\vec{k}}^{(w)}>$ in the denominator of $I_1$ can be expressed in the term of radial distribution function(RDF)$g_{_{lw}}$, reading
\begin{align}
\label{IID11}
<R_{\vec{k}}^{(l)}R_{-\vec{k}}^{(w)}>=\delta_{l,w}+A\int_0^{\infty} dr~ r \sin(k r) (g_{_{lw}}(r)-1)
\end{align}
with $A=4\pi\sqrt{\rho^{(l)}\rho^{(w)}}/k$. In the coefficient $A$, $\rho^{(l)}$ is the averaged number density $<\rho^{(l)}(\vec{x})>$ of particles of the $l$-th species and obtained by $N_l$ dividing the total volume of the system. $\rho^{(w)}$ has the similar meaning of $\rho^{(l)}$, and is for the $w$-th species. $\delta_{l,w}$ is the Kronecker delta function. The RDF $g_{_{lw}}$ is defined as $g_{_{lw}}(r)=<\sum_{n=1}^{N_l}\sum_{m=1}^{N_w} \delta(\vec{r}+\vec{x}_{n}^{(l)}-\vec{x}_{m}^{(w)})>/(N_l\rho_{w})$. The MD approach to $g_{_{lw}}(r)$ is described in the following. For a configuration of the system obtained by MD, we take a particle of the $l$-th species at the center of a spherical shell. The shell has the radius of $r$ and a thickness of $\Delta r$. The volume of the shell is noted by $\Delta V$. Then we count the number of all the particles of the $w$-th species in the shell. The counting number is divided by the product of the density $\rho^{(w)}$ and the volume $\Delta V$. The result after the division then is averaged over all the particles of the $l$-th species, leading to $g_{_{lw}}(r)$.\\

By borrowing the idea of the RDF, the numerator $<\mathcal{T}_{\vec{k}}^{(l)}\mathcal{T}_{-\vec{k}}^{(w)}>$ in $I_1$ can be expressed as
\begin{align}
\label{IID12}
<\mathcal{T}_{\vec{k}}^{(l)}\mathcal{T}_{-\vec{k}}^{(w)}>=&\delta_{w,l}<\sum_{n=1}^{N_l} (\hat{k}\cdot\ddot{\vec{x}}_n^{(l)})^2>/N_l\nonumber\\
&+A\int dr~r \sin(kr)h_{_{lw}}(r)
\end{align}
with $h_{_{lw}}(r)=<\sum_{n=1}^{N_l}\sum_{m=1}^{N_w} (\hat{k}\cdot\ddot{\vec{x}}^{(l)}_n) (\hat{k}\cdot\ddot{\vec{x}}^{(w)}_m)\delta(\vec{r}+\vec{x}_n^{(l)}-\vec{x}_m^{(w)})>/(N_l\rho^{(w)})$. Here, the double dots on $\vec{x}^{(l)}_n$ mean the acceleration $\ddot{\vec{x}}^{(l)}_n$ of the $n$-th particle of the $l$-th species. The function $h_{_{lw}}(r)$ is referred to as radial distribution function of force (RDFF), and can be obtained by MD, similar to what we have done for $g_{_{lw}}(r)$. For a given configuration by MD, we take a particle of the $l$-th species at the center with a spherical shell around the particle. The shell is with the radius $r$ and the volume $\Delta V$. Then we sum the accelerations of all the particles of the $w$-th species in the shell and take the component of the sum only along $\hat{k}$ direction. We make a product of the component and the acceleration of the center particle also along $\hat{k}$ direction. The product then is $\sum_{m=1}^{N_w} (\hat{k}\cdot\ddot{\vec{x}}^{(l)}_n) (\hat{k}\cdot\ddot{\vec{x}}^{(w)}_m)$. The product is divided by the $\rho^{(w)}\Delta V$ to get a quantity. This quantity is averaged by all the particles of the $l$-th species, resulting in $h_{_{lw}}(r)$.\\

The derivations for Eq.(\ref{IID11}) and Eq.(\ref{IID12}) have been put in Appendix B. 

\subsection{factor $I_2$}
\label{IIE}
In the numerator of $I_2$, we have the definition of $\mathcal{F}_{\vec{k}}^{(l)}=\sum_{p}(i\hat{k}\cdot F_{\vec{k}}^{(l,p)})$. The factor $F_{\vec{k}}^{(l,p)}$ is the Fourier component of the force of a particle of the $l$-th species acted by a particle of the $p$-th species, which can be found in Eq.(\ref{IIA4}). Now we reformulate $\mathcal{F}_{\vec{k}}^{(l)}$ to be
\begin{align}
\label{IIE13}
\mathcal{F}_{\vec{k}}^{(l)}=\sum_{p}\frac{1}{V_{\vec{k}}}\int  B_{\vec{k},\vec{k}_1}^{(l,p)} \rho_{\vec{k}-\vec{k}_1}^{(l)}\rho_{\vec{k}_1}^{(p)}d \vec{k}_1.
\end{align}
Here, $B_{\vec{k},\vec{k}_1}^{(l,p)}$ is originated from the Fourier transformation on the potential of the pairwise interaction between a particle of the $l$-th species and a particle of the $p$-th species. The coefficient $B_{\vec{k},\vec{k}_1}^{(l,p)}$ is functional of the species and the wave vectors. $V_{\vec{k}}$ is the volume for the integration in the reciprocal space, which will be canceled later in the calculation. The details for Eq.(\ref{IIE13}) can be found in Appendix C.\\

Based on Eq.(\ref{IIE13}), the force correlation function in the numerator of $I_2$ is time dependent and reads
\begin{align}
\label{IIE14}
<\mathcal{F}_{\vec{k}}^{(l)}\mathcal{T}_{-\vec{k}}^{(w)}>=\frac{1}{(V_{\vec{k}})^2}\sum_{p,q}\int  d \vec{k}_1d \vec{k}_2~<B_{\vec{k},\vec{k}_1}^{(l,p)}B_{-\vec{k},\vec{k}_2}^{(w,q)}>\mathcal{J}
\end{align}
with $\mathcal{J}=<\rho_{\vec{k}-\vec{k}_1}^{(l)}\rho_{\vec{k}_1}^{(p)} R_{-\vec{k}-\vec{k}_2}^{(w)}R_{\vec{k}_2}^{(q)}>.$
Here, we have decoupled the factor $<B_{\vec{k},\vec{k}_1}^{(l,p)}B_{-\vec{k},\vec{k}_2}^{(w,q)}>$ from $\mathcal{J}$ because they are not strongly related. Due to the translational invariance of the system, $\vec{k}_2=-\vec{k}_1$ must be held to conserve the momentum. To go further, we reformulate Eq.(\ref{IIE14}) to be
\begin{align}
\label{IIE15}
<\mathcal{F}_{\vec{k}}^{(l)}\mathcal{T}_{-\vec{k}}^{(w)}>=\frac{1}{(V_{\vec{k}})^2}\sum_{p,q}\int  d \vec{k}_1~\mathcal{J}_1\cdot \mathcal{J}_2
\end{align}  
with 
\begin{align}
&\mathcal{J}_1=<B_{\vec{k},\vec{k}_1}^{(l,p)}B_{-\vec{k},-\vec{k}_1}^{(w,q)}><R_{\vec{k}-\vec{k}_1}^{(l)}R_{\vec{k}_1}^{(p)} R_{-\vec{k}+\vec{k}_1}^{(w)}R_{-\vec{k}_1}^{(q)}>,\nonumber\\
&\mathcal{J}_2=\frac{<\rho_{\vec{k}-\vec{k}_1}^{(l)}\rho_{\vec{k}_1}^{(p)} R_{-\vec{k}+\vec{k}_1}^{(w)}R_{-\vec{k}_1}^{(q)}>}{<R_{\vec{k}-\vec{k}_1}^{(l)}R_{\vec{k}_1}^{(p)} R_{-\vec{k}+\vec{k}_1}^{(w)}R_{-\vec{k}_1}^{(q)}>}.\nonumber 
\end{align}
We apply the Wick theorem to both the numerator and the denominator of $\mathcal{J}_2$. Considering the momentum conservation for correlation functions, we have $\mathcal{J}_2=\mathcal{G}_{\vec{k}-\vec{k}_1}^{(l,w)}\mathcal{G}_{\vec{k}_1}^{(p,q)}$.\\

By using Eq.(\ref{IIE15}), the denominator in $I_2$ is
\begin{align}
\label{IIE16}
<\mathcal{T}_{\vec{k}}^{(l)}\mathcal{T}_{-\vec{k}}^{(w)}>=\frac{1}{(V_{\vec{k}})^2}\sum_{p,q}\int  d \vec{k}_1~\mathcal{J}_1
\end{align}  
because of $\mathcal{J}_2=1$ at the initial time. Note that Eq.(\ref{IIE16}) is different from Eq.(\ref{IID12}) in the expression for the same quantity $<\mathcal{T}_{\vec{k}}^{(l)}\mathcal{T}_{-\vec{k}}^{(w)}>$. This is because we understand the same quantity from different physical views. By using Eq.(\ref{IIE15}) and Eq.(\ref{IIE16}), we cancel $1/(V_{\vec{k}})^2$ in $I_2$. For convenience, we define a coefficient
\begin{align}
\label{IIE17}
\beta_{\vec{k},\vec{k}_1}^{(l,w,p,q)}=\frac{\mathcal{J}_1}{\sum_{p,q}\int  d \vec{k}_1~\mathcal{J}_1},
\end{align}
to get
\begin{align}
\label{IIE18}
I_2=\sum_{p,q}\int  d \vec{k}_1~\beta_{\vec{k},\vec{k}_1}^{(l,w,p,q)}\cdot \mathcal{G}_{\vec{k}-\vec{k}_1}^{(l,w)}\mathcal{G}_{\vec{k}_1}^{(p,q)}.
\end{align}
Putting all the information mentioned above together, we rewrite Eq.(\ref{IIC10}) as
\begin{align}
\label{IIE19}
&\frac{d^4 \mathcal{G}_{\vec{k}}^{(l,w)}}{dt^4}+\zeta_1\cdot \frac{d^2 \mathcal{G}_{\vec{k}}^{(l,w)}}{dt^2}+\zeta_2\cdot \mathcal{G}_{\vec{k}}^{(l,w)}\nonumber\\
&=I_1\cdot \left[\sum_{p,q}\int  d \vec{k}_1~\beta_{\vec{k},\vec{k}_1}^{(l,w,p,q)}\cdot \mathcal{G}_{\vec{k}-\vec{k}_1}^{(l,w)}\mathcal{G}_{\vec{k}_1}^{(p,q)}\right].
\end{align}
In the following, we introduce the MD approach to the coefficient $\beta_{\vec{k},\vec{k}_1}^{(l,w,p,q)}$.\\

\subsection{coefficient $\beta_{\vec{k},\vec{k}_1}^{(l,w,p,q)}$ }
\label{IIF}
In order to solve $\beta_{\vec{k},\vec{k}_1}^{(l,w,p,q)}$, we need calculate $\mathcal{J}_1$ firstly as the numerator and then integrate $\mathcal{J}_1$ over the whole reciprocal space as the denominator according to the definition Eq.(\ref{IIE17}). To solve $\mathcal{J}_1$, we will use two different expressions of $\mathcal{T}_{\vec{k}}^{(l)}$ as the bridge to connect $\mathcal{J}_1$ and MD simulations.\\

Suppose we can express $\mathcal{T}_{\vec{k}}^{(l)}=\sum_{p}\frac{1}{V_{\vec{k}}}\int  \mathcal{I}^{(l,p)}_{\vec{k}, \vec{k}_1}d \vec{k}_1$ in the reciprocal space. we compare the above expression to Eq.(\ref{IIE13}) and have 
\begin{align}
\label{IIF20}
\mathcal{I}^{(l,p)}_{\vec{k}, \vec{k}_1}&=  B_{\vec{k},\vec{k}_1}^{(l,p)} R_{\vec{k}-\vec{k}_1}^{(l)}R_{\vec{k}_1}^{(p)}\nonumber\\
&=\frac{i}{\sqrt{N_l}} \sum_{n,m} e^{i\vec{k}\cdot \vec{x}_n^{(l)}}(\hat{k}\cdot  y^{(l,p)}_{n,\vec{k}_1})e^{i\vec{k}_1\cdot \vec{x}_m^{(p)}}
\end{align}
for each component at one given wave vector $\vec{k}_1$. The first line in the above equation is from Eq.(\ref{IIE13}) and the second line is from the definition of $\mathcal{T}_{\vec{k}}^{(l)}$ expressed in the term of accelerations. For the second line in Eq.(\ref{IIF20}), it has
$y^{(l,p)}_{n,\vec{k}_1}=\sum_m \ddot{\vec{x}}^{(l,p)}_{n,m} e^{-i\vec{k}_1\cdot \vec{x}_m^{(p)}}$, in which $\ddot{\vec{x}}^{(l,p)}_{n,m}$ is the acceleration component of the $n$-th particle of the $l$-th species subjected to the $m$-th particle of the $p$-th species only. Since the two expressions in the two lines reveal the same physical meaning at the same wave vector, they must be equivalent to each other. All the acceleration components can be calculated by MD. Thus, it is no problem to calculate $y^{(l,p)}_{n,\vec{k}_1}$ and $\mathcal{I}^{(l,p)}_{\vec{k}, \vec{k}_1}$ further. Here, it is unnecessary to calculate the coefficient $B_{\vec{k},\vec{k}_1}^{(l,p)}$ alone. Finally, we calculate $\mathcal{J}_1$ by $\mathcal{J}_1=<\mathcal{I}^{(l,p)}_{\vec{k}, \vec{k}_1}\mathcal{I}^{(w,q)}_{-\vec{k}, -\vec{k}_1}>$ through the MD simulations. The details for Eq.(\ref{IIF20}) can be found in Appendix D.\\

In this study, we focus on the case of $l=w$. In this case, $p=q$ must be held and we have
\begin{align}
\label{IIF21}
\mathcal{J}_1=\frac{N_p}{N_l}\left<\sum_{n,u}(\hat{k}\cdot  \ddot{\vec{x}}^{(l,p)}_{n,u} )^2R_{\vec{k}_1}^{(p)}R_{-\vec{k}_1}^{(p)}\right>,
\end{align}
which also can be found in Appendix D. For convenience, the notations are simplified by replacing $\beta_{\vec{k},\vec{k}_1}^{(l,l,p,p)}$ with $\beta_{\vec{k},\vec{k}_1}^{(l,p)}$ and replacing $\mathcal{G}_{\vec{k}}^{(l,l)}$ with $\mathcal{G}_{\vec{k}}^{(l)}$. Eq.(\ref{IIE19}) then is rewritten as
\begin{align}
\label{IIF22}
&\frac{d^4 \mathcal{G}_{\vec{k}}^{(l)}}{dt^4}+\frac{2}{ \alpha^{(l)}}\frac{d^2 \mathcal{G}_{\vec{k}}^{(l)}}{dt^2}+\frac{1}{ [\alpha^{(l)}]^2}\mathcal{G}_{\vec{k}}^{(l)}\nonumber\\
&=\frac{1}{\eta^2k^2}\frac{<\mathcal{T}_{\vec{k}}^{(l)}\mathcal{T}_{-\vec{k}}^{(l)}>}{<R_{\vec{k}}^{(l)}R_{-\vec{k}}^{(l)}>}\cdot \left[\sum_{p}\int  d \vec{k}_1~\beta_{\vec{k},\vec{k}_1}^{(l,p)}\cdot \mathcal{G}_{\vec{k}-\vec{k}_1}^{(l)}\mathcal{G}_{\vec{k}_1}^{(p)}\right].
\end{align}

\subsection{initial conditions}
\label{IIG}
To solve Eq.(\ref{IIE19}), we need four initial conditions $\mathcal{G}_{\vec{k},0}^{(l,w)}$, $\dot{\mathcal{G}}_{\vec{k},0}^{(l,w)}$, $\ddot{\mathcal{G}}_{\vec{k},0}^{(l,w)}$ and $\dddot{\mathcal{G}}_{\vec{k},0}^{(l,w)}$. We use $0$ in the subscript of a quantity to represent the value of the quantity at the initial time $t=0$. For example, $\dot{\mathcal{G}}_{\vec{k},0}^{(l,w)}$ is the value of $\dot{\mathcal{G}}_{\vec{k}}^{(l,w)}$ at $t=0$. According to the definition $\mathcal{G}_{\vec{k}}^{(l,w)}=<\rho_{\vec{k}}^{(l)}R_{-\vec{k}}^{(w)}>/<R_{\vec{k}}^{(l)}R_{-\vec{k}}^{(w)}>$, we have $\mathcal{G}_{\vec{k},0}^{(l,w)}=1$. Due to the no correlation of current and density, $\dot{\mathcal{G}}_{\vec{k},0}^{(l,w)}=0$ is obtained. By using Eq.(\ref{IIA6}), the initial value of $\ddot{\mathcal{G}}_{\vec{k}}^{(l,w)}$ can be obtained from the right hand side of Eq.(\ref{IIA6}), leading to $\ddot{\mathcal{G}}_{\vec{k},0}^{(l,w)}=-\delta_{w,l}/[\alpha^{(l)}<R_{\vec{k}}^{(l)}R_{-\vec{k}}^{(w)}>]$, which can be found in Appendix E. Finally, $\dddot{\mathcal{G}}_{\vec{k},0}^{(l,w)}$ is the value of $-<\ddot{R}_{\vec{k}}^{(l)}\dot{R}_{-\vec{k}}^{(w)}>/<R_{\vec{k}}^{(l)}R_{-\vec{k}}^{(w)}>$ by using the property of the Liouville operator in Eq.(\ref{IIB7}). Due to the no correlation of $\ddot{R}_{\vec{k}}^{(l)}$ and $\dot{R}_{-\vec{k}}^{(w)}$ at the initial time, we have $\dddot{\mathcal{G}}_{\vec{k},0}^{(l,w)}=0$. We list the four initial conditions here 
\begin{align}
\label{IIG23}
&\mathcal{G}_{\vec{k},0}^{(l,w)}=1,~~~~~~~~~~~~~~~~~~~~~~~~~~~~\dot{\mathcal{G}}_{\vec{k},0}^{(l,w)}=0,\nonumber\\
&\ddot{\mathcal{G}}_{\vec{k},0}^{(l,w)}=-\frac{1}{ \alpha^{(l)}}\frac{\delta_{w,l}}{<R_{\vec{k}}^{(l)}R_{-\vec{k}}^{(w)}>},~~~~\dddot{\mathcal{G}}_{\vec{k},0}^{(l,w)}=0,
\end{align}
for clarity.\\

Starting from the initial time, we can expand $\mathcal{G}_{\vec{k}}^{(l)}$ after a time step $\Delta t$ by using the Taylor expansion. We have
\begin{align}
\label{IIG24}
\mathcal{G}_{\vec{k}}^{(l)}(\Delta t)\approxeq 1-\frac{1}{2!}\frac{1}{ \alpha^{(l)}}\frac{1}{<R_{\vec{k}}^{(l)}R_{-\vec{k}}^{(l)}>}(\Delta t)^2
\end{align}
showing that $\mathcal{G}_{\vec{k}}^{(l)}$ decays from its initial value. The full behavior of $\mathcal{G}_{\vec{k}}^{(l)}$ should be solved from Eq.(\ref{IIF22}) combined with the initial conditions of Eq.(\ref{IIG23}). Due to the nonlinearity, it is  difficult to solve Eq.(\ref{IIF22}) numerically. In this study, we are only interested in the static value of $\mathcal{G}_{\vec{k}}^{(l)}$ when time approaches infinity. Fortunately, the static value of $\mathcal{G}_{\vec{k}}^{(l)}$ can be obtained analytically and then is used to judge if the glass relaxation is ergodic or non-ergodic.

\section{results}
Eq.(\ref{IIF22}) is general and can be used for a glass comprised of multi-species. In this study, we take a binary LJ glass comprised of two species as an example. We follow the model of the binary LJ glass in Ref.(\onlinecite{Kob1}), which is successful for MD simulations. The two species in the system are noted by A and B respectively. The LJ potential has been given in Eq.(\ref{IIA3}). The LJ parameters and the masses of the particles can be found in Ref.(\onlinecite{Kob1}). We only extend the simulation size by increasing the number of A particles to 6400 and the number of B particles to 1600. Lammps is used for the MD simulations~\cite{Lmp}. The initial temperature is started at $\eta=5$ to melt the system and then is decreased to the intended temperature for equilibrium. The details for the MD simulation can be found in Ref.(\onlinecite{Kob1}).\\

\subsection{density correlation}
\label{IIIA}
The RDFs at various temperatures are presented in Fig. 1. In the figure, the RDFs have been shifted for clarity. The definition of RDFs can be found in Eq.(\ref{IID11}). Fig.1(a) is for the RDFs of A particles, noted by $g_{AA}$, showing that the system has the feathers of liquid at the temperatures $\eta\geq 1$. When the temperature is decreased lower than $\eta =0.6$, the second peak of RDFs begins to split into two small peaks, which is the feature of glass~\cite{Zallen}. Fig.1(b) is for the partial RDFs $g_{AB}$, which is similar to $g_{AA}$. The main difference between $g_{AA}$ and $g_{AB}$ is that the first peak in $g_{AB}$ sifts to a lower value. This is because the radius of B particles is smaller than that of A particles~\cite{Kob1}. Fig.1(c) is $g_{BB}$ for B particles. The number density of B particles is smaller than that of A particles. Therefore, the probability of finding a B particle in the first peak of Fig.1(c) is lower than the probability of the first peak in Fig.1(a) and (b). The split of the second peak is still observed in Fig.1(c).\\

\begin{figure}[!h]
\includegraphics[width=0.4\textwidth]{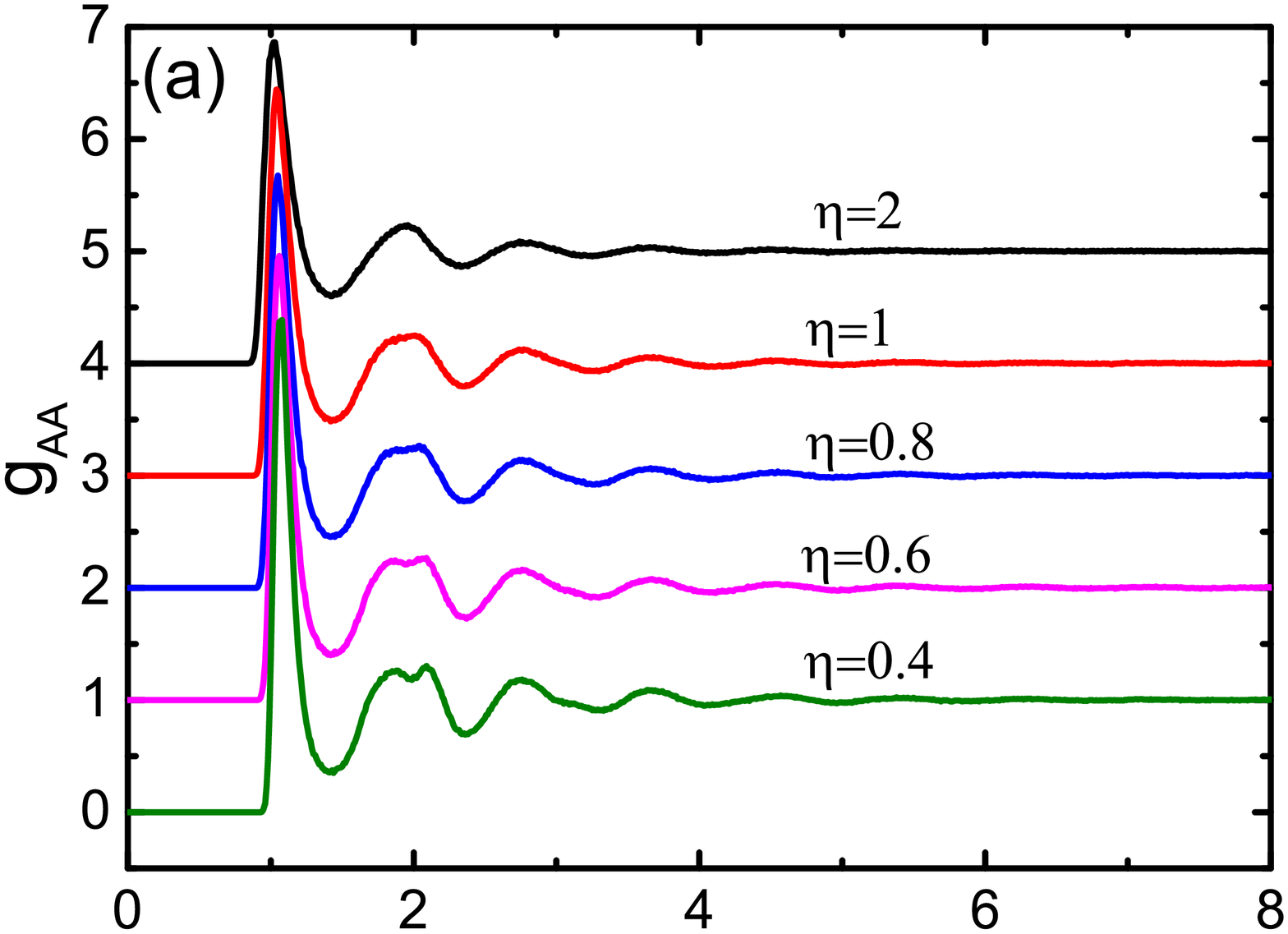}\\[0.5cm]
\includegraphics[width=0.4\textwidth]{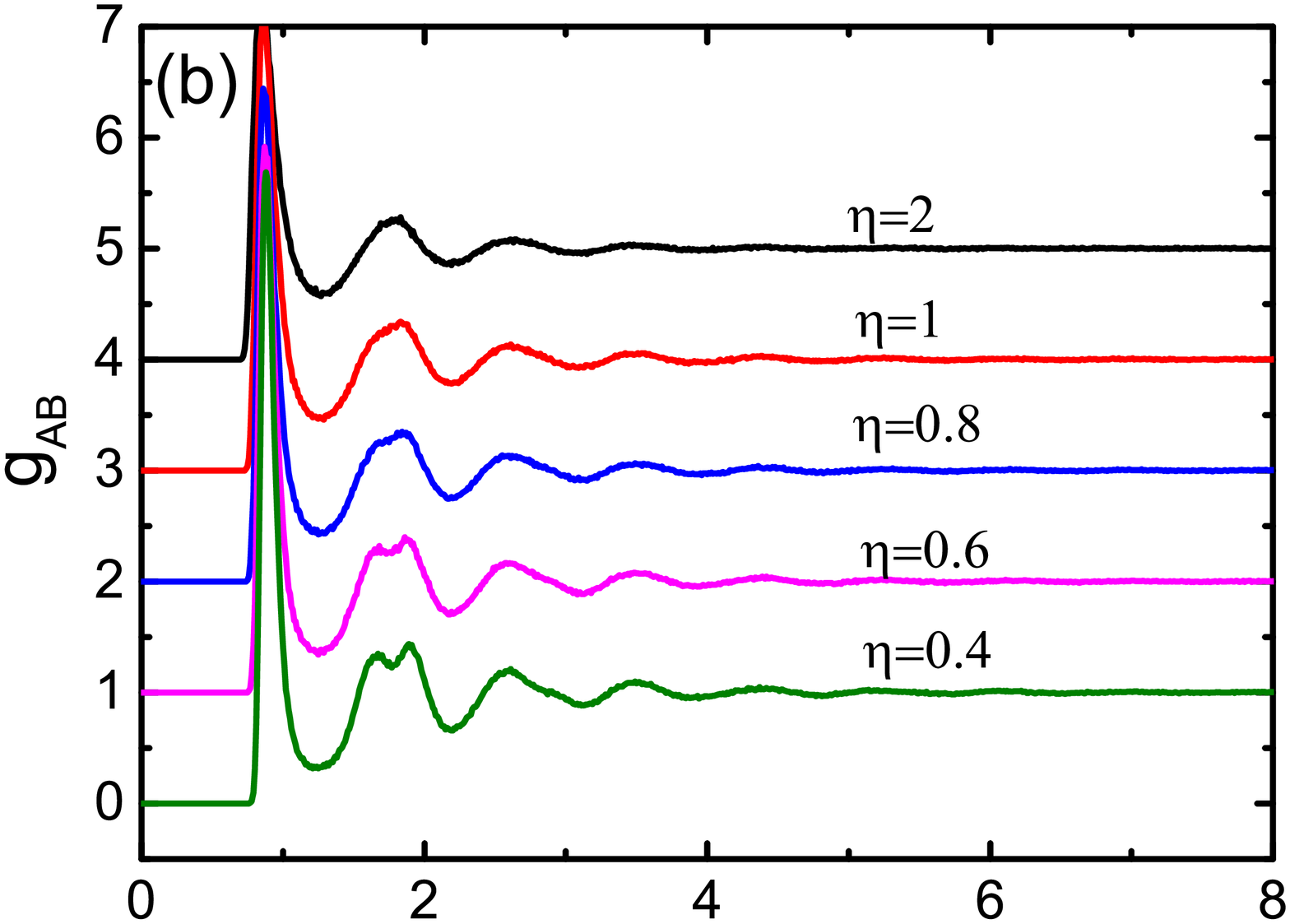}\\[0.5cm]
\includegraphics[width=0.4\textwidth]{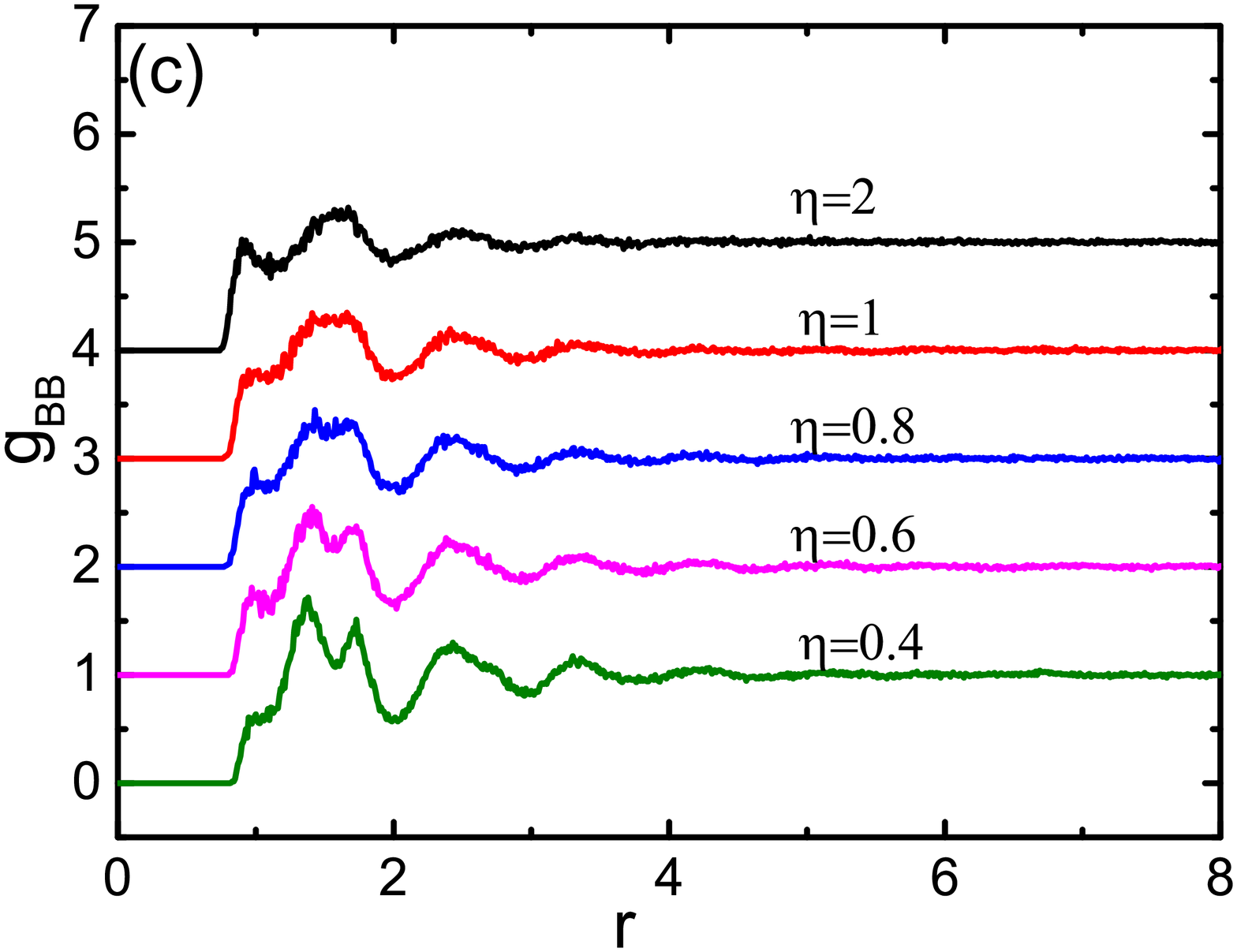}
\caption{\label{fig1}Radial distribution functions (RDFs). (a) $g_{AA}$ is the RDFs for A particles. (b) $g_{AB}$ is the partial RDFs of B particles around a particle of A species. $g_{BA}$ is equivalent to $g_{AB}$ and does not repeat in the figure. (c) $g_{BB}$ is the RDFs for B particles.}
\end{figure}

The RDF of glass is insensitive to temperatures, and is not a proper parameter to reflect the structural transition for the glass. Some other order parameters have been proposed by many authors for the structural transition~\cite{Bailey,Tong}. In this study, we suggest a new parameter based on the force fluctuation for the purpose, which will be shown later.\\

By using the results of RDF, the DCFs can be obtained through Eq.(\ref{IID11}) and results are shown in Fig. 2. We use $d_{lw}$ to represent the DCF $<R^{(l)}_{\vec{k}}R^{(w)}_{-\vec{k}}>$ for short notation. Data of $d_{lw}$ have been shifted for clarity in Fig. 2. It indicates that the DCFs vary in the range of small wave vectors, say $k<15$, but go uniformly for large $k$. $d_{AA}$ and $d_{BB}$ go to $1$ while $d_{AB}$ goes to zero. The nonzero of $d_{AA}$ and $d_{BB}$ at large $k$ means that the local relaxations play an important role in the whole glass relaxation and can not be neglected.
\begin{figure}[!h]
\includegraphics[width=0.4\textwidth]{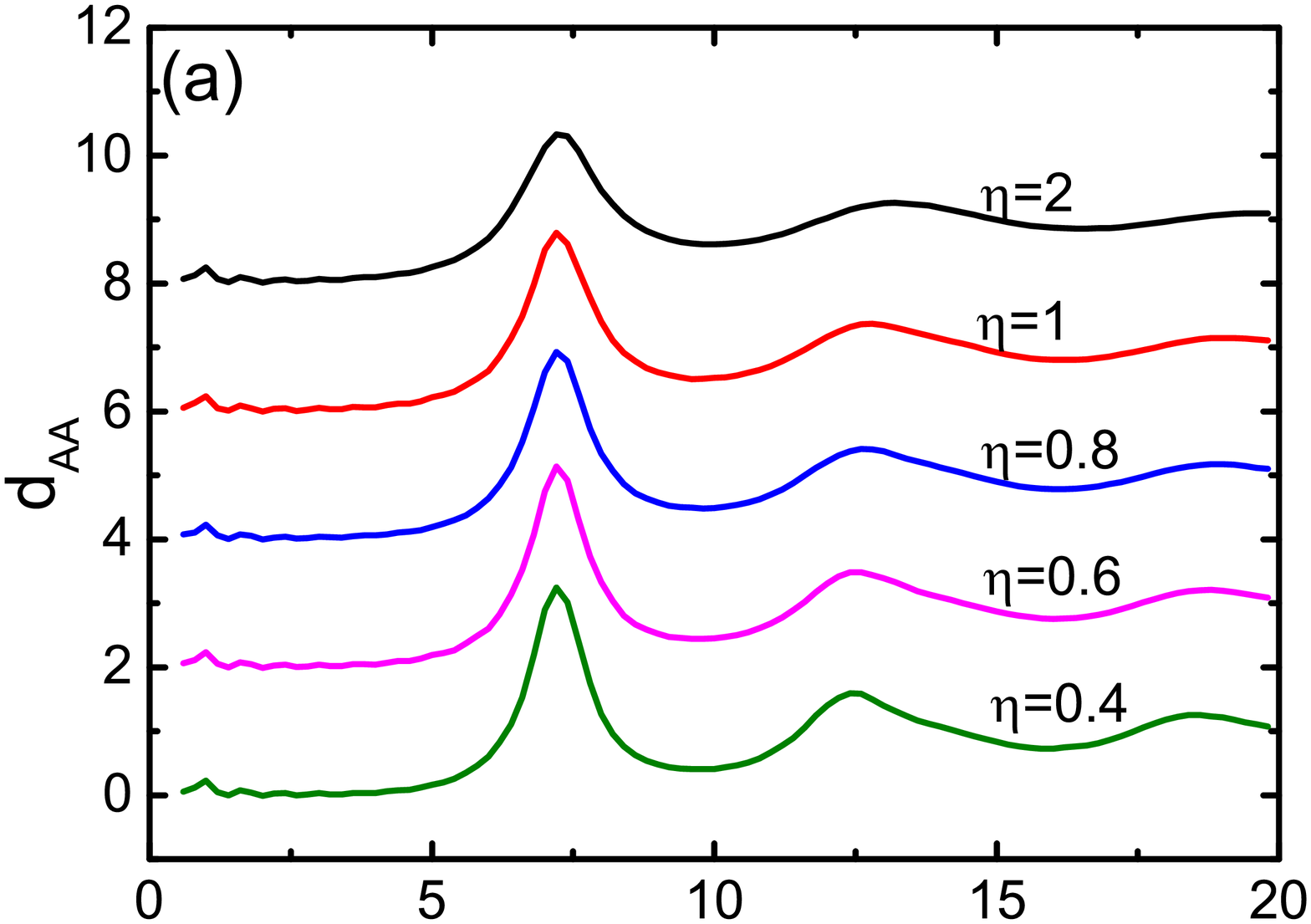}\\[0.5cm]
\includegraphics[width=0.4\textwidth]{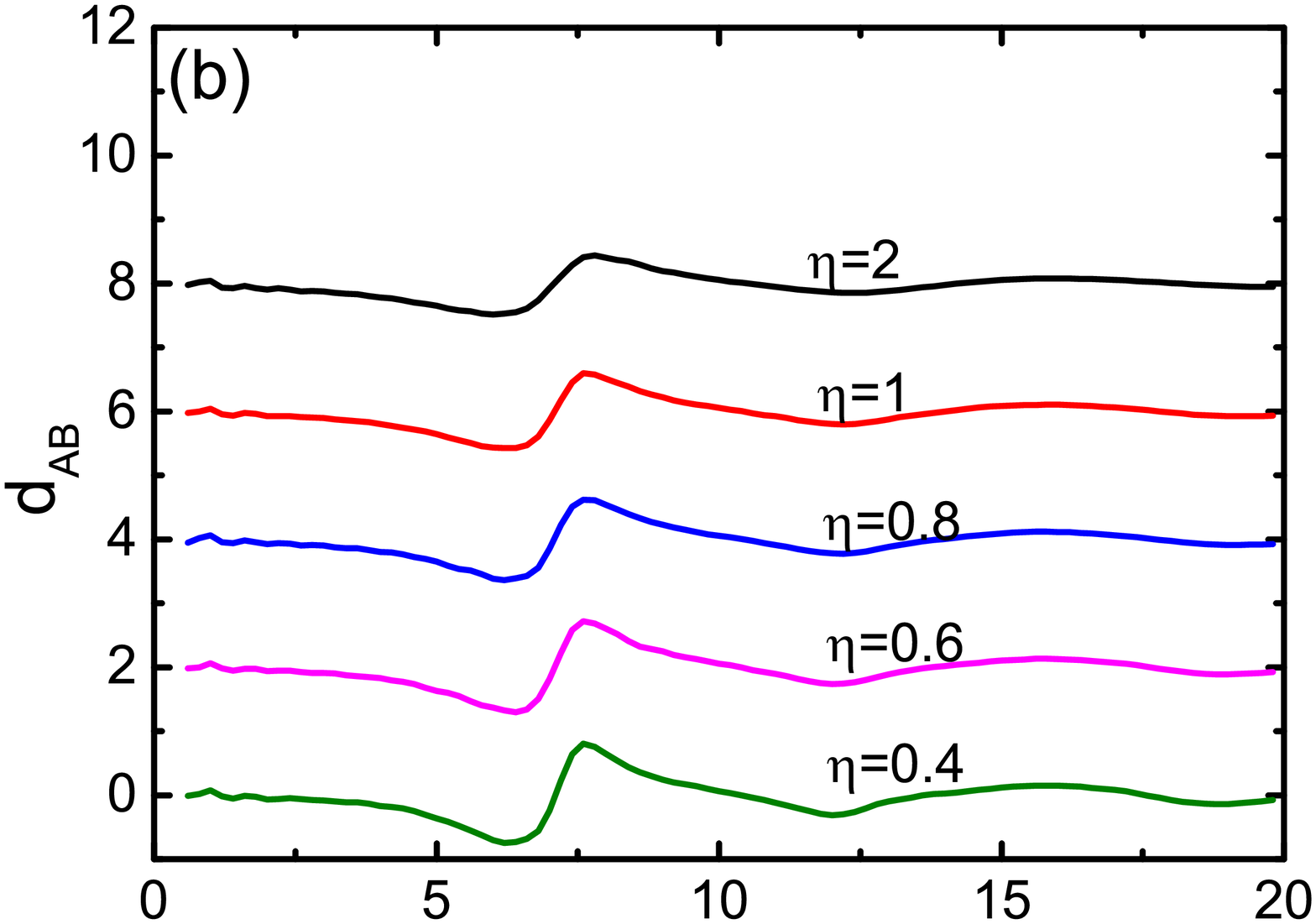}\\[0.5cm]
\includegraphics[width=0.4\textwidth]{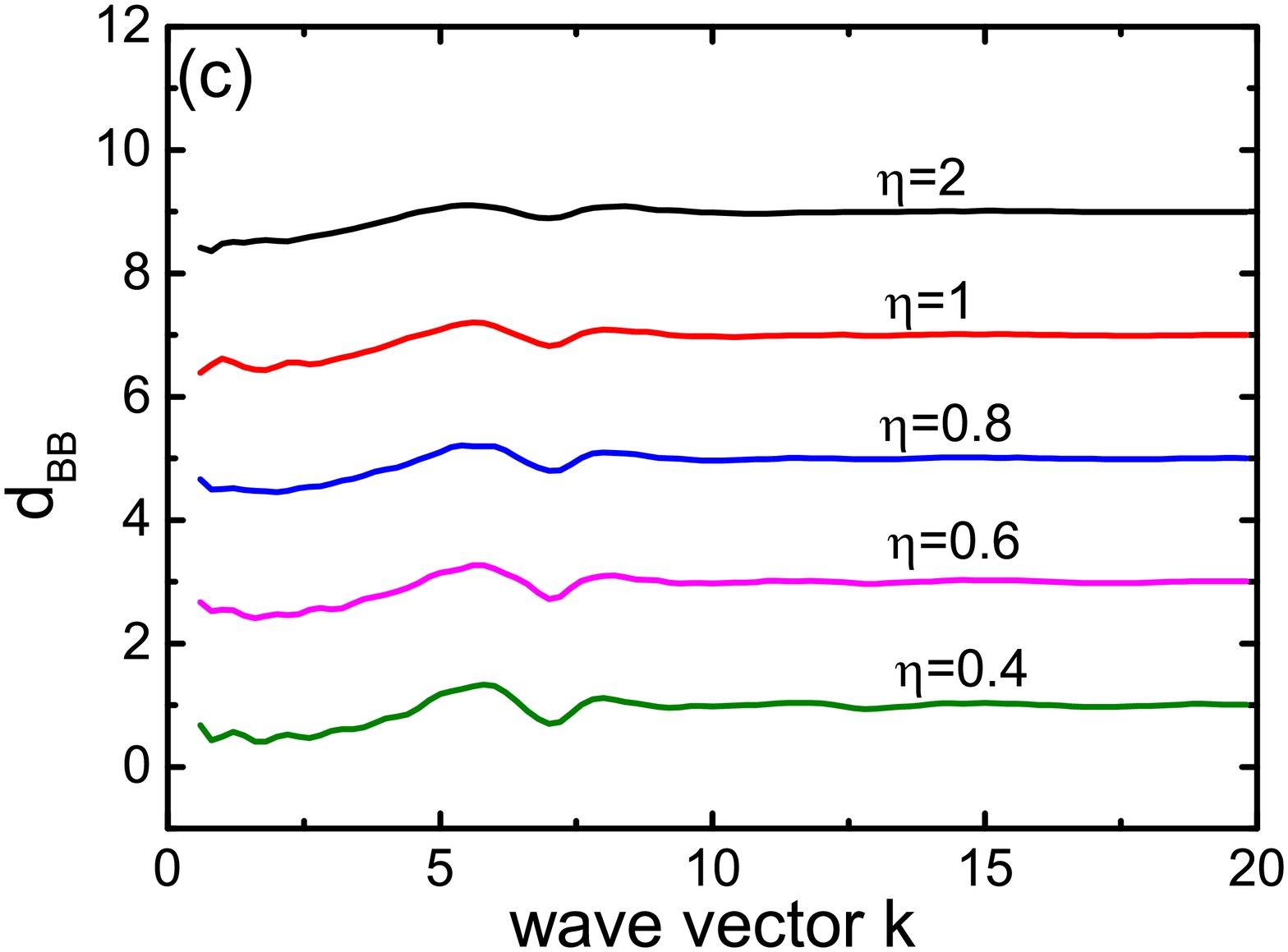}
\caption{\label{fig2} Density correlation functions(DCFs). (a) $d_{AA}$ is the DCFs for $<R_{\vec{k}}^{(A)}R_{-\vec{k}}^{(A)}>$. (b) $d_{AB}$ is the DCFs for $<R_{\vec{k}}^{(A)}R_{-\vec{k}}^{(B)}>$. (c) $d_{BB}$ is the DCFs  for $<R_{\vec{k}}^{(B)}R_{-\vec{k}}^{(B)}>$. }
\end{figure}

\subsection{force correlation}
\label{IIIB}
The RDFF $h_{lw}$ defined in Eq.(\ref{IID12}) is shown in Fig.3, in which data have been shifted for clarity. 
\begin{figure}[!h]
\includegraphics[width=0.4\textwidth]{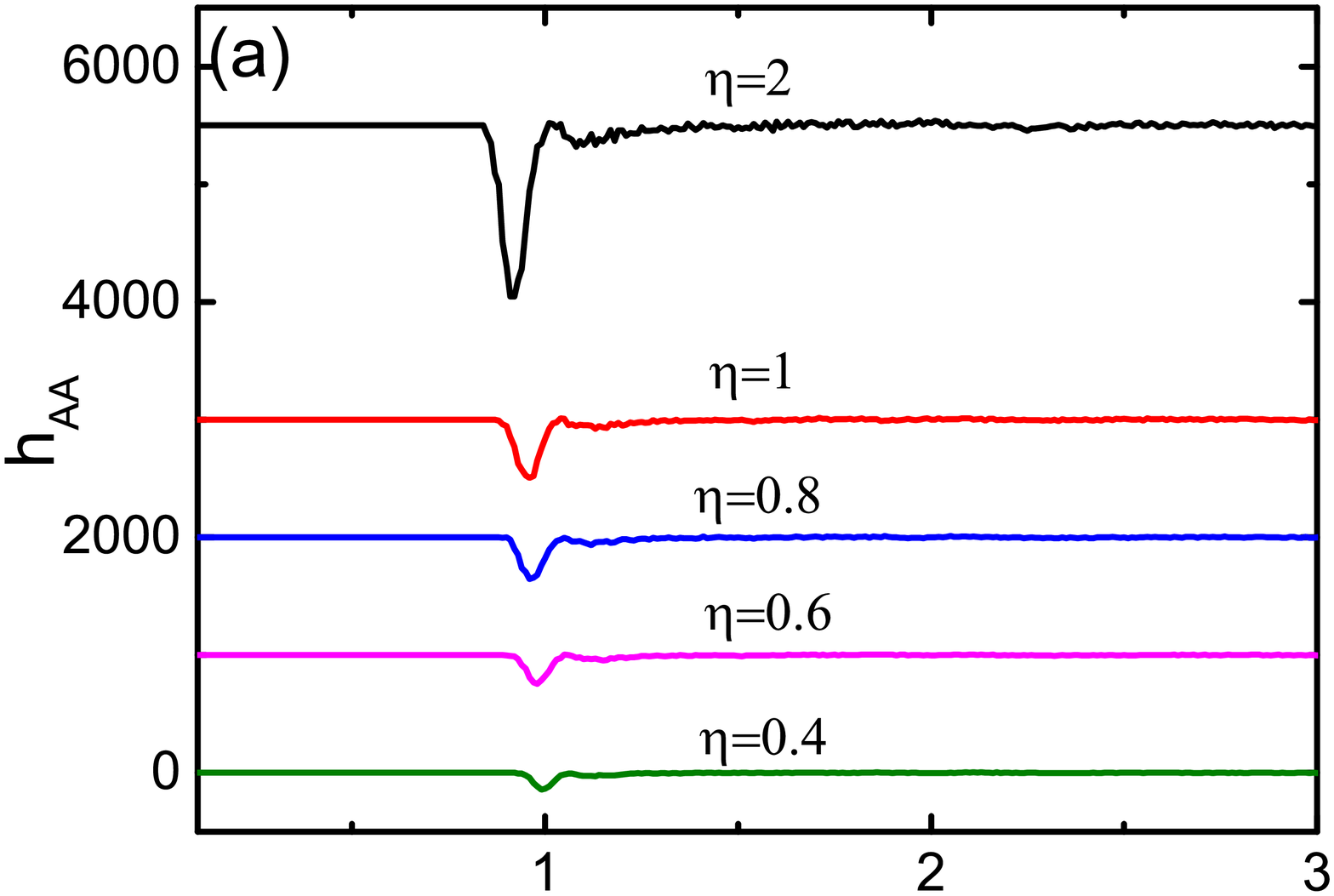}\\[0.5cm]
\includegraphics[width=0.4\textwidth]{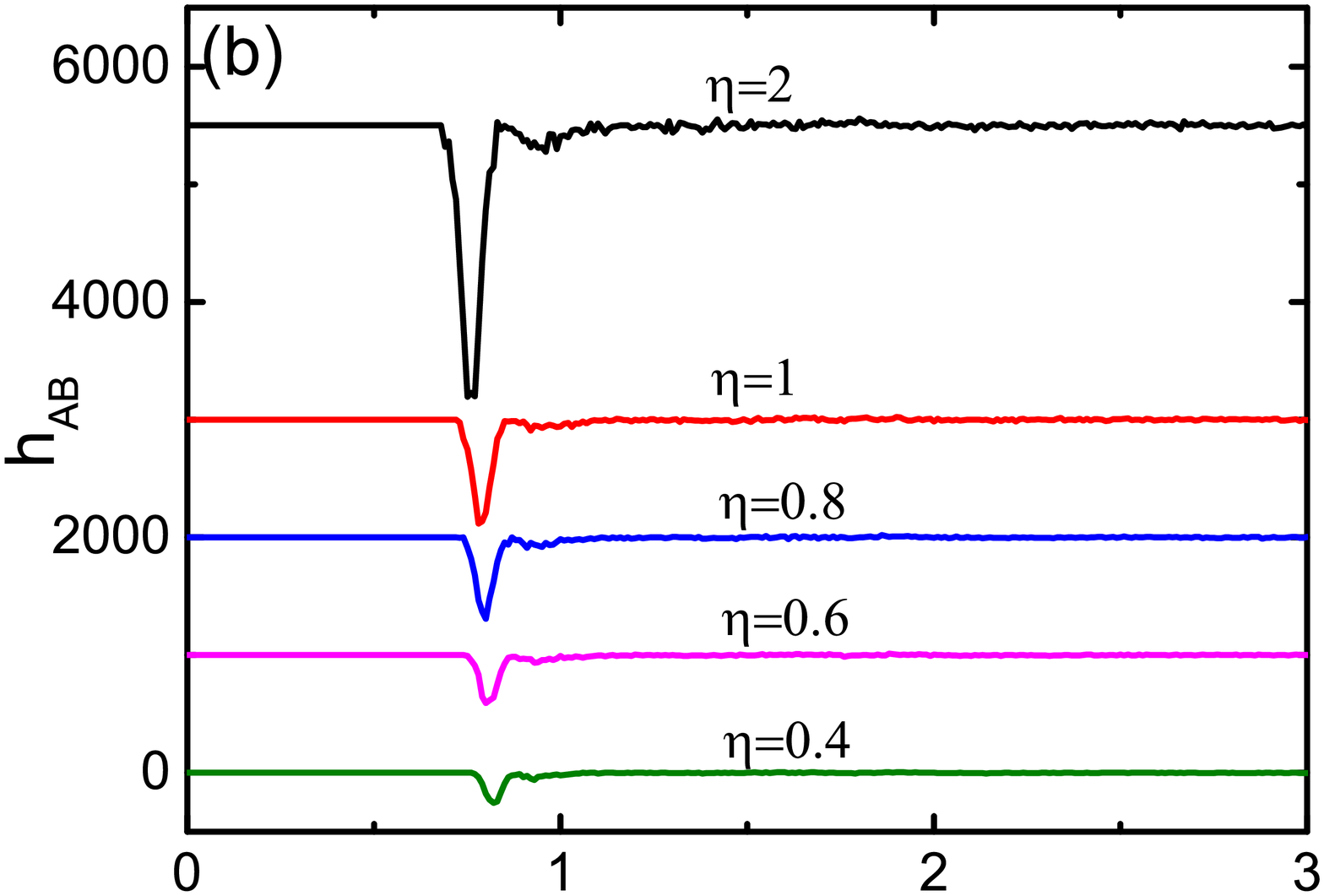}\\[0.5cm]
\includegraphics[width=0.4\textwidth]{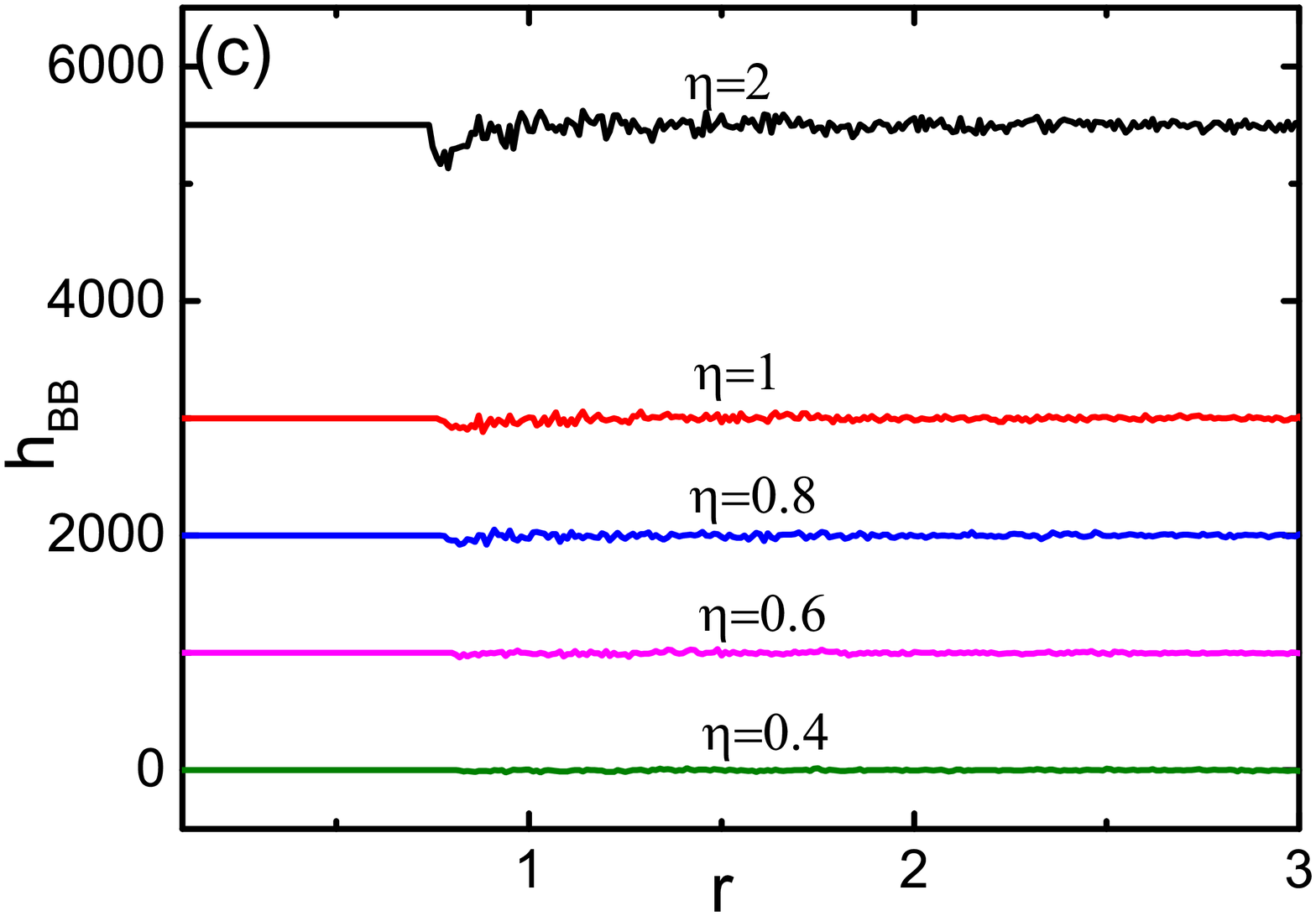}
\caption{\label{fig3} Radial distribution function of force(RDFF).(a) $h_{AA}$ is the RDFF for A particles. (b) $h_{AB}$ is the RDFF of B particles around a particle of A species. $h_{BA}$ is equivalent to $h_{AB}$ and does not repeat in the figure. (c) $h_{BB}$ is the RDFF for B particles.}
\end{figure}
The functions are not positive because the action and anti action are opposite in direction. There exists only one valley for each RDFF. The width of the valleys is very narrow and close to 0.1, meaning that the particles interacting with the center particle almost concentrate in a shell. The thickness of the shells is small corresponding to the narrow width of the valleys. The positions of the valleys are insensitive to $\eta$ as is figured out from Fig.3. Thus, the averaged distance for the interaction between particles is almost not changed with $\eta$. Such result is coincident to the results of RDF in Fig.1. At a high temperature such as $\eta=2$, the magnitude of RDFF is large and decreased with the decreasing of the temperature. Due to the small number density of the B particles, the average distance between the B particles is very large. Almost no action between B particles can be detected, which has been revealed in Fig.3(c). \\

According to the RDF results in Fig.1, the number density of particles in the shell interacting with the center particle is not changed with the temperature. Thus, the number of particles in the shell is fixed for the various temperatures. But why does the magnitude of the RDFF decrease with the decrease of $\eta$? This is because the RDFF is determined by the distribution of the particles in the shell. Suppose the particles in the shell distribute uniformly. The forces of the particles in the shell acting on the center particle are balanced, leading to zero of the RDFF. Once the particles in the shell deviate from the balanced configuration, the force acting on the center particle is nonzero. The anti action on the particles in the shell by the center particle is also nonzero. In this way, the RDFF will be nonzero. When $\eta$ is high, particles in the shell have large momentum and move intensively. There exists a high probability of the particles in the shell deviating from the balanced configuration. When cooling down, the particles seek configurations with the minimum force acting on the particles. In this way, the magnitude of the RDFF decreases with the temperature.\\

The degree of the deviation from the balanced configuration for the particles in the shell can be reflected by the force acting on the center particle. Since the net force of the system is averaged to be zero, we define the force fluctuation $\lambda_{l}=<\sum_n(\hat{k}\cdot \ddot{\vec{x}}^{(l)}_n)^2>/N_l$ of an individual particle of the $l$-th species to show the degree of the deviation. The force fluctuation can be defined for A species with $l=A$ or B species with $l=B$. Based on the results of RDFF, the force fluctuations should be strongly related to the temperature $\eta$, which has been proved in Fig.4. 
\begin{figure}[!h]
\includegraphics[width=0.4\textwidth]{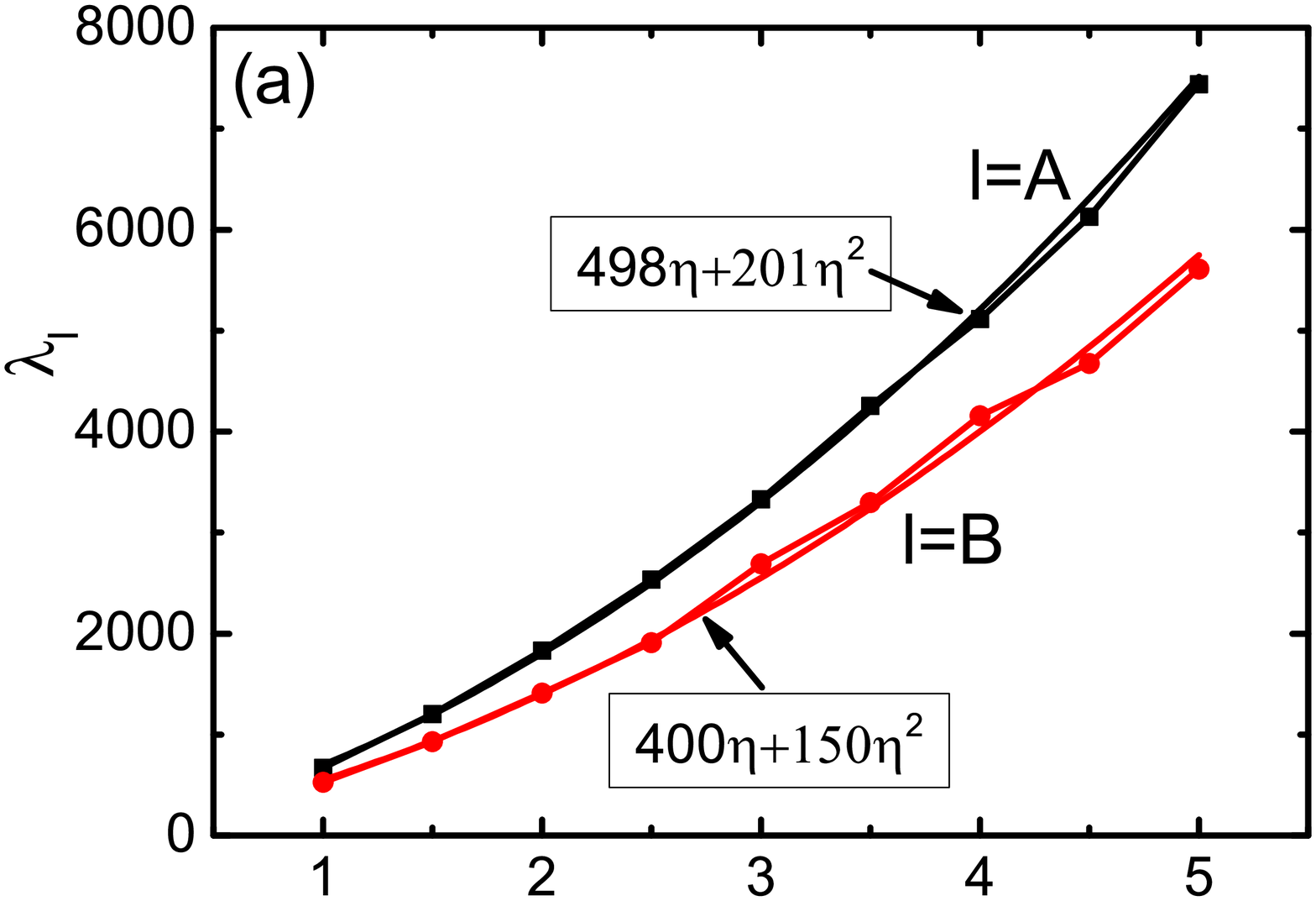}\\[0.5cm]
\includegraphics[width=0.4\textwidth]{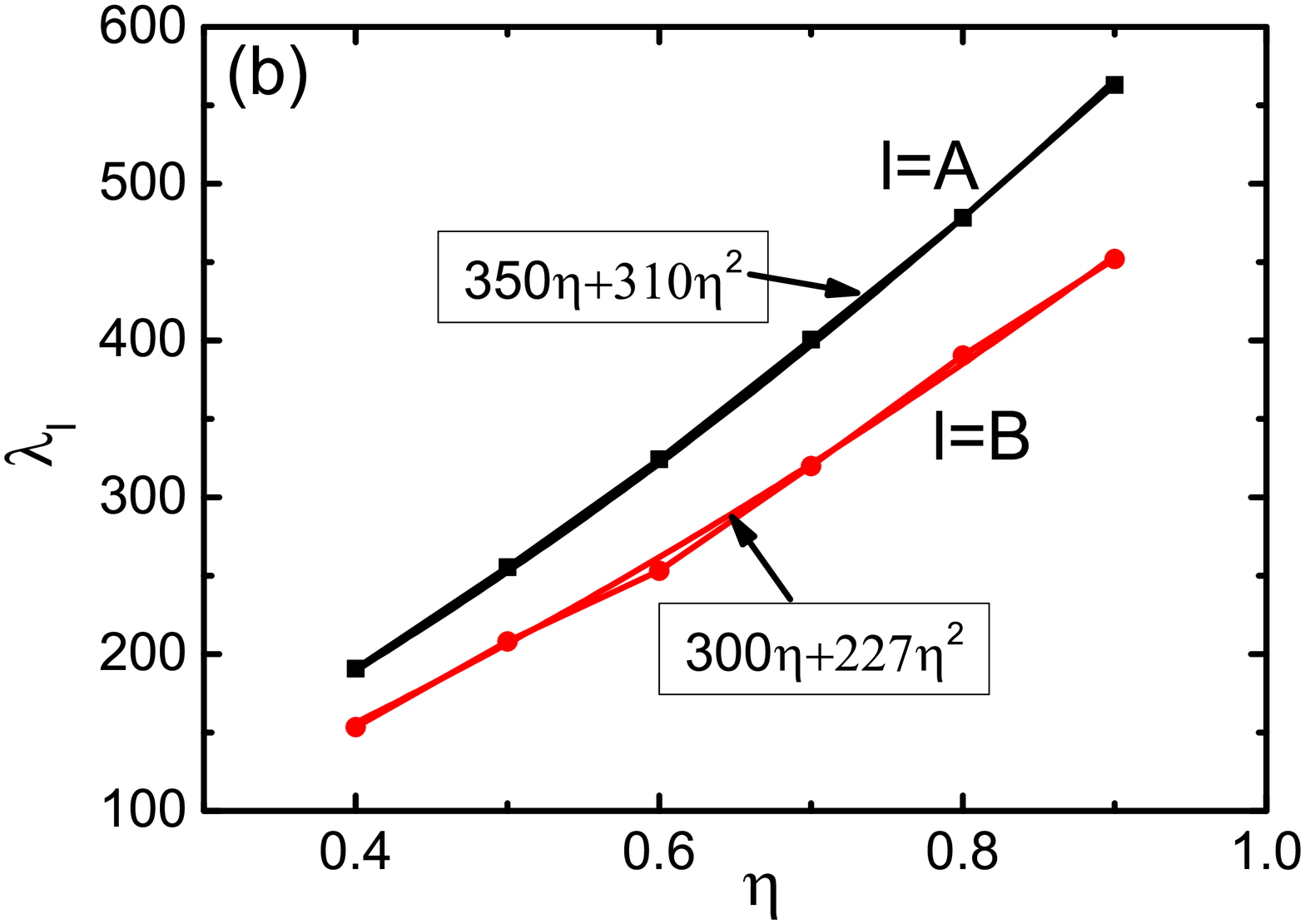}
\caption{\label{fig4} Force fluctuation of one individual particle defined by $\lambda_{l}=<\sum_n(\hat{k}\cdot \ddot{\vec{x}}^{(l)}_n)^2>/N_l$.(a) $\lambda_l$ is functional of $\eta$ for the section of $\eta\ge 1$.(b) $\lambda_l$ is functional of $\eta$ for the section of $\eta < 1$.}
\end{figure}
To show the behaviors of the $\eta$ dependence, we fit the plot for two sections. One is for $\eta\ge 1$ in Fig.4(a) and the other is for $\eta<1$ in Fig.4(b). It shows that the plot can be fit by $\lambda_l\approxeq \xi_1 \eta+\xi_2\eta^2$ with the coefficients $\xi_1$ and $\xi_2$ indicated in the figure. We think the main effect of $\eta$ on the structural transition is to change the degree of force unbalance in the shells. Thus, we suggest $\lambda_l$ to be an order parameter for structural transition. However, in this study, we focus on the ergodicity of the glass relaxation. The relation between the force fluctuation and structural transition is not included here.\\

To show the force fluctuation non-locally, the force correlation function(FCF) $f_{lw}=<\mathcal{T}_{\vec{k}}^{(l)}\mathcal{T}_{-\vec{k}}^{(w)}>$ is used with the wave vector $\vec{k}$ introduced. The function $f_{lw}$ can be obtained by the MD simulations through Eq.(\ref{IID12}). Results are presented in Fig. 5. Data in the figure are not shifted.
\begin{figure}[!h]
\includegraphics[width=0.4\textwidth]{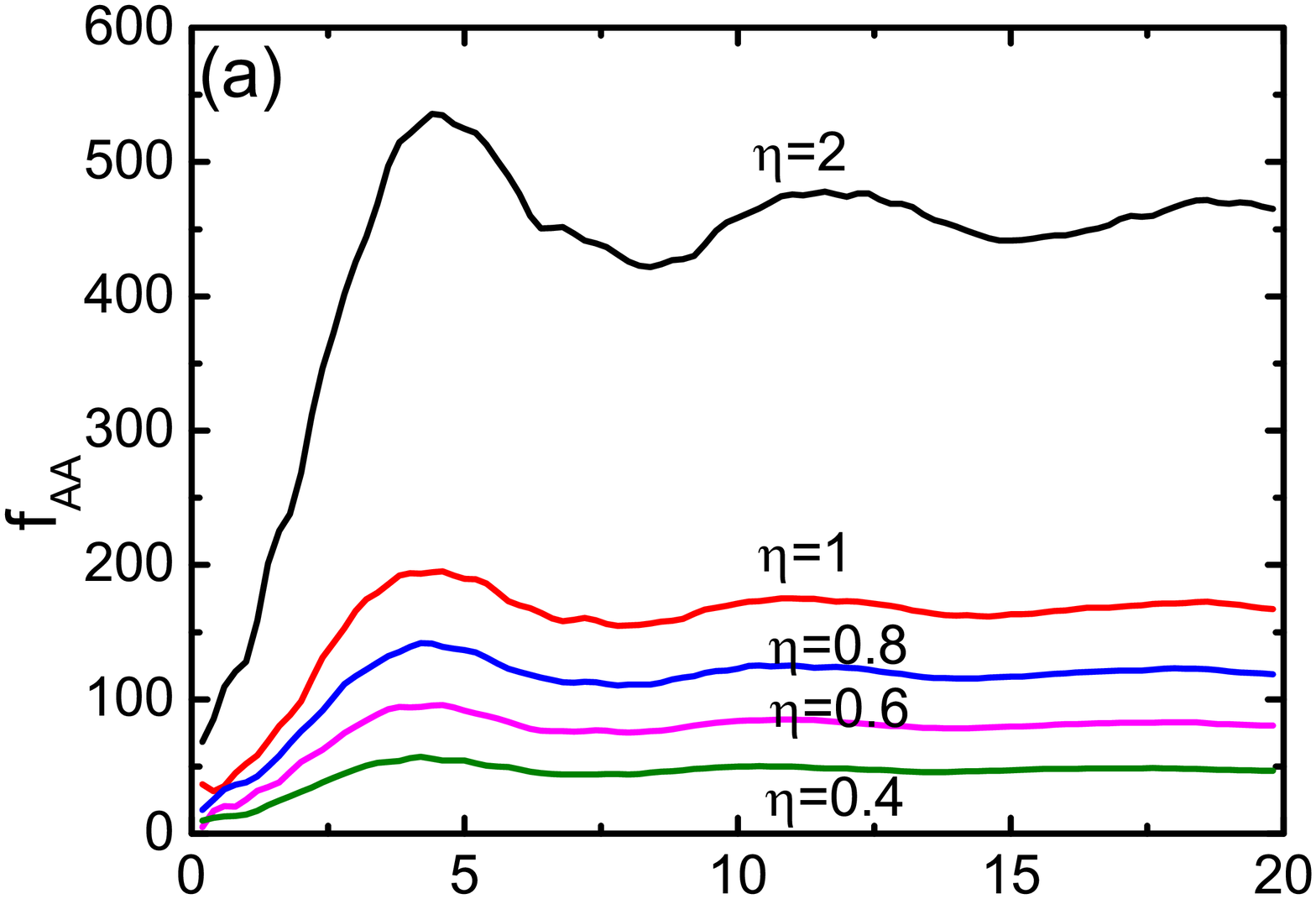}\\[0.5cm]
\includegraphics[width=0.4\textwidth]{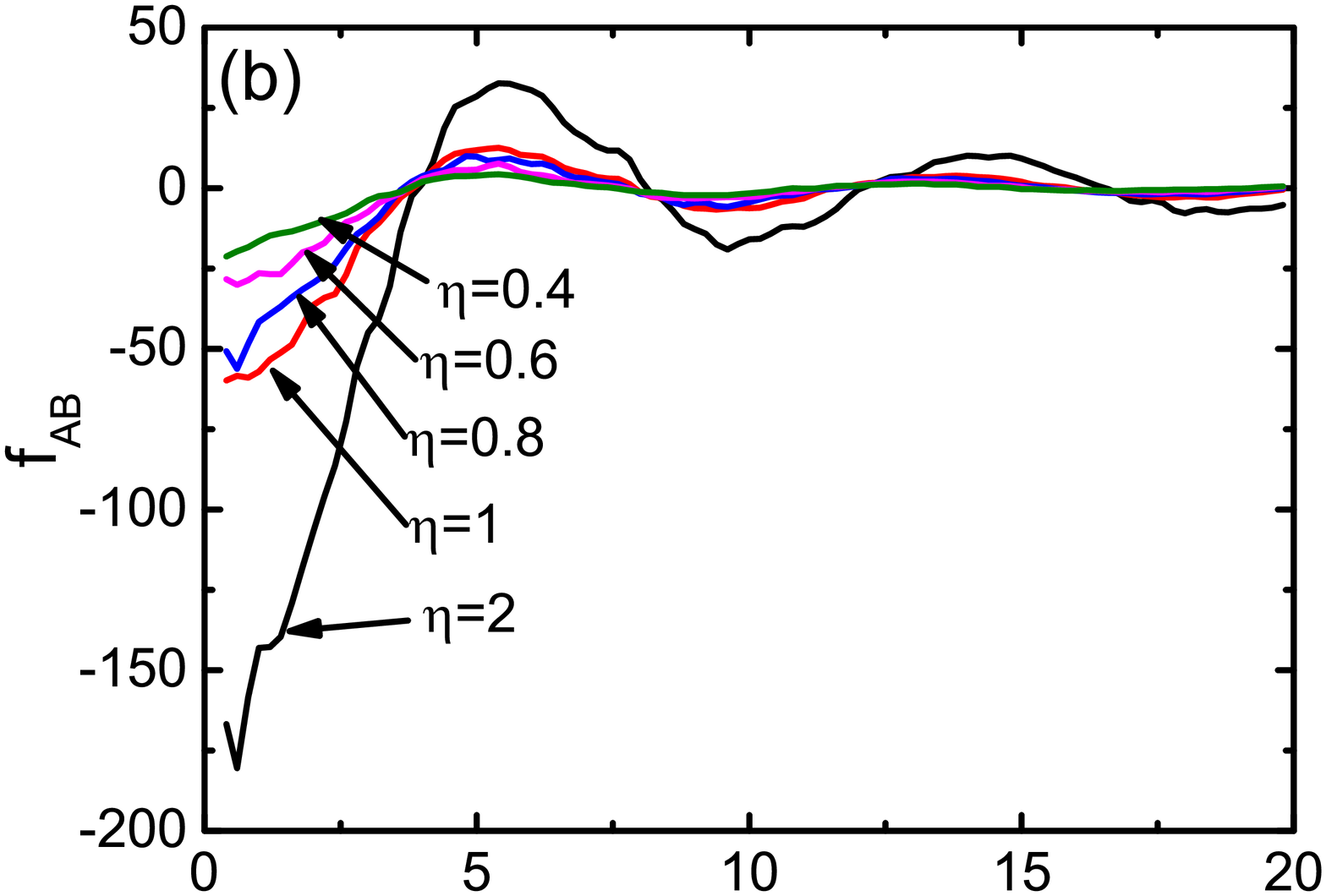}\\[0.5cm]
\includegraphics[width=0.4\textwidth]{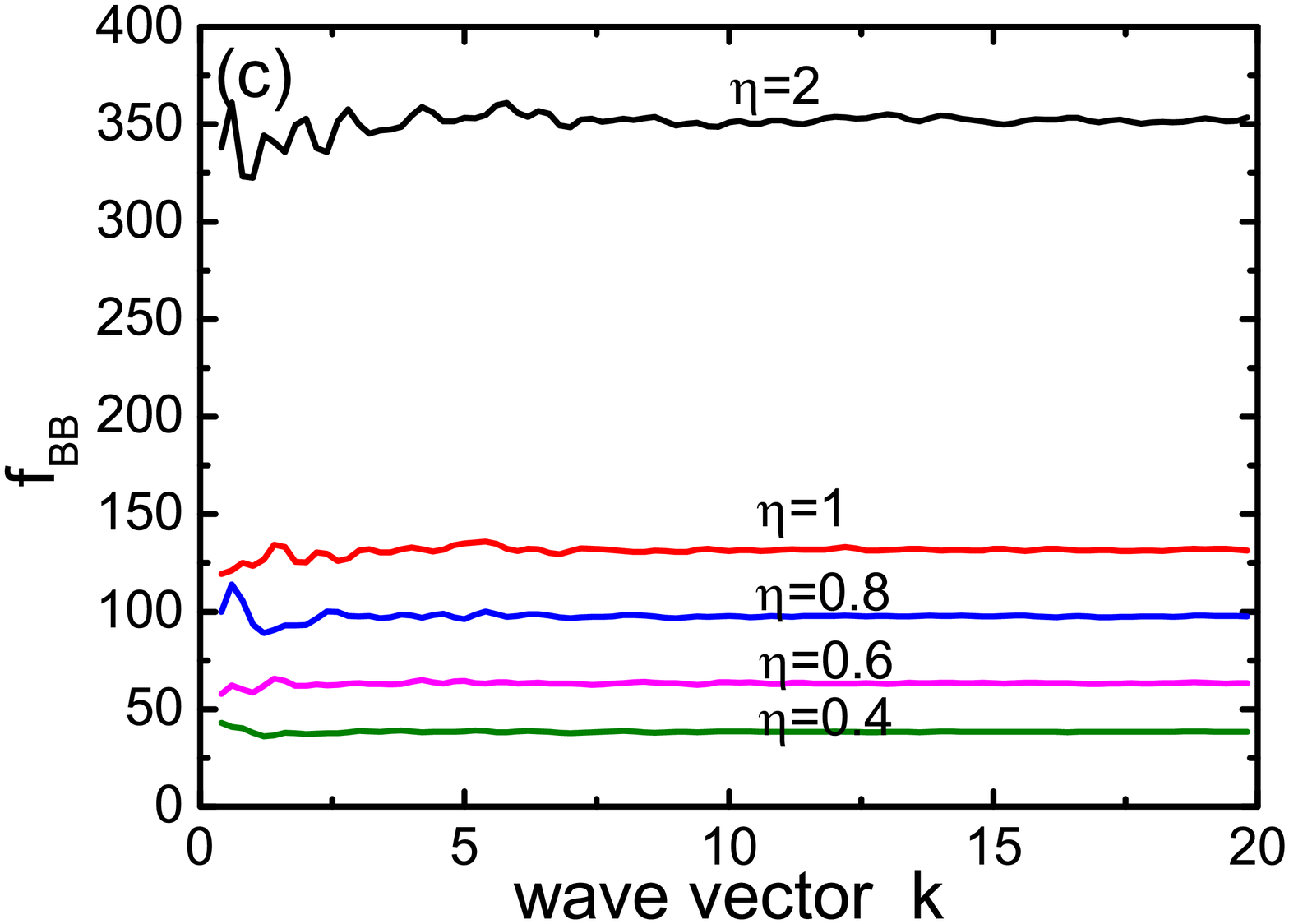}
\caption{\label{fig5} Force correlation functions(FCFs).(a)$f_{AA}$ is the FCFs for $<\mathcal{T}_{\vec{k}}^{(A)}\mathcal{T}_{-\vec{k}}^{(A)}>$.(b)$f_{AB}$ is the FCFs for $<\mathcal{T}_{\vec{k}}^{(A)}\mathcal{T}_{-\vec{k}}^{(B)}>$.(c)$f_{BB}$ is the FCFs for $<\mathcal{T}_{\vec{k}}^{(B)}\mathcal{T}_{-\vec{k}}^{(B)}>$.}
\end{figure}
Principally, at $k=0$, all the FCFs should be zero. However, it is difficult to be caught by the numerical calculation at $k=0$ because of the singularity. With the increasing of $k$, the size we are interested in the system is decreased by $1/k$. For large $k$, the function $f_{AA}$ is dominated by the force fluctuation $\lambda_l$ with $l=A$, and $f_{BB}$ is with $l=B$. In Fig.5(a) and (c), the FCF of A species or B species is functional of $\eta$ at any wave vector. However, in Fig.5(b), the FCF for the correlation between the A species and the B species is around zero for the glass. Thus, in Eq.(\ref{IIE19}), the solution for the correlation of A and B particles is $\mathcal{G}^{(A,B)}_{\vec{k}}=0$ due to the approximation of $f_{AB}=0$ in the right hand side of the equation. That means $\mathcal{G}^{(A,B)}_{\vec{k}}=0$ contributes little to $\mathcal{G}^{(A,A)}_{\vec{k}}$ and $\mathcal{G}^{(B,B)}_{\vec{k}}$. Such conclusion is consistent to Eq.(\ref{IIF22}), where no $\mathcal{G}^{(A,B)}_{\vec{k}}$ appears on the right hand side. The interaction between the A particles and the B particles has been included in the coefficients $\beta^{(l,p)}_{\vec{k},\vec{k}_1}$ for the solutions of $\mathcal{G}^{(A)}_{\vec{k}}$ and $\mathcal{G}^{(B)}_{\vec{k}}$ in Eq.(\ref{IIF22}).\\

Our results are obtained for the temperature $\eta>0.4$. For the temperature lower than 0.4, it needs more time for the MD simulations to equilibrate the configurations. The results obtained for $\eta <0.4$ are not reliable in this study. Generally, we can assume $<\mathcal{T}_{\vec{k}}^{(l)}\mathcal{T}_{-\vec{k}}^{(w)}>\approxeq \sum_n \xi_n\eta^n$ with $\xi_n$ functional of $\vec{k}$ and $n\ge 1$ for the glass.\\

\subsection{ergodicity}
\label{IIIC}
Eq.(\ref{IIG24}) tells us that the normalized DCF $\mathcal{G}_{\vec{k}}^{(l,w)}$ decays from the initial time. Due to the nonlinearity, it is difficult to solve the equation for the full behaviors of $\mathcal{G}_{\vec{k}}^{(l,w)}$. In this work, we focus on the static result of Eq.(\ref{IIF22}). For this purpose, our equation can be simplified by dropping off the first two derivative terms since the static result is independent on time. Then, Eq.(\ref{IIF22}) reads
\begin{align}
\label{IIIC25}
\frac{\eta^2k^2}{[\alpha^{(l)}]^2}\mathcal{G}_{\vec{k}}^{(l)}=\frac{<\mathcal{T}_{\vec{k}}^{(l)}\mathcal{T}_{-\vec{k}}^{(l)}>}{<R_{\vec{k}}^{(l)}R_{-\vec{k}}^{(l)}>}\sum_{p}\int  d \vec{k}_1~\beta_{\vec{k},\vec{k}_1}^{(l,p)}\cdot \mathcal{G}_{\vec{k}-\vec{k}_1}^{(l)}\mathcal{G}_{\vec{k}_1}^{(p)}
\end{align}
with the factor $\eta^2k^2$ moved to the left hand side for convenience. \\

In the glass, $\eta$ is lower than $1$. In the limit case, $\eta$ approaches zero. The denominator $<R_{\vec{k}}^{(l)}R_{-\vec{k}}^{(l)}>$ is nearly independent on the temperature according to Fig.(\ref{fig2}). The numerator $<\mathcal{T}_{\vec{k}}^{(l)}\mathcal{T}_{-\vec{k}}^{(l)}>$ has been fitted by $\sum_n\xi_n\eta^n$ in Sec.(\ref{IIIB}). Thus, we can drop off one $\eta$ on both sides of Eq.(\ref{IIIC25}) and then let $\eta$ approach zero. Finally only one term remains on the right hand side of Eq.(\ref{IIIC25}) as
\begin{align}
\label{IIIC26}
0=\frac{\xi_1}{<R_{\vec{k}}^{(l)}R_{-\vec{k}}^{(l)}>}\sum_{p}\int  d \vec{k}_1~\beta_{\vec{k},\vec{k}_1}^{(l,p)}\cdot \mathcal{G}_{\vec{k}-\vec{k}_1}^{(l)}\mathcal{G}_{\vec{k}_1}^{(p)}.
\end{align}
In the above equation, the coefficient $\beta_{\vec{k},\vec{k}_1}^{(l,p)}$ is defined as $\mathcal{J}_1/\int d\vec{k} \mathcal{J}_1$ with $\mathcal{J}_1=\left<\sum_{n,u}(\hat{k}\cdot  \ddot{\vec{x}}^{(l,p)}_{n,u} )^2R_{\vec{k}_1}^{(p)}R_{-\vec{k}_1}^{(p)}\right>N_p/N_l$ from Eq.(\ref{IIF21}). In $\mathcal{J}_1$, the factor $(\hat{k}\cdot  \ddot{\vec{x}}^{(l,p)}_{n,u} )^2$ is  positive and $R_{\vec{k}_1}^{(p)}$ is conjugate to $R_{-\vec{k}_1}^{(p)}$, which makes $R_{\vec{k}_1}^{(p)}R_{-\vec{k}_1}^{(p)}$ positive. That means $\mathcal{J}_1$ is always positive and  $\beta_{\vec{k},\vec{k}_1}^{(l,p)}$ is always positive too. $\beta_{\vec{k},\vec{k}_1}^{(l,p)}$ is smaller than 1. In $\beta_{\vec{k},\vec{k}_1}^{(l,p)}$, the common factors $\eta$ in the numerator and the denominator are canceled, and then $\beta_{\vec{k},\vec{k}_1}^{(l,p)}$ is less dependent on the temperature $\eta$. Considering $1\ge \mathcal{G}^{(l)}_{\vec{k}}\ge 0$, we get the static solution to Eq.(\ref{IIIC26}), which is $\mathcal{G}^{(l)}_{\vec{k}}=0$ for all the wave vector $\vec{k}$ and all the species $l$. Note that the above result is obtained under the condition of $\eta\approxeq 0$. For $\eta>0$, the static result $\mathcal{G}^{(l)}_{\vec{k}}=0$ must be held since the higher temperature moves the particles much more intensively and relax the system in a shorter time. For the case of $\xi_1=0$, $\eta^2$ is dropped off on the both sides of Eq.(\ref{IIIC25}), followed by setting $\eta \approxeq 0$. We still obtain the solution of $\mathcal{G}^{(l)}_{\vec{k}}=0$.\\

The solution $\mathcal{G}^{(l)}_{\vec{k}}=0$ to Eq.(\ref{IIIC26}) indicates that the overlap between the final configuration of the glass and the initial configuration is averaged to be zero. Starting from the initial configuration of the glass, the glass will experience all the possible configurations in the relaxation. Thus, the glass relaxation is ergodic. 

\section{conclusion}
In this study, we have derived an equation for the glass relaxation. In the derivation, we don't use the Zwanzig-Mori projection method, which makes our equation is different from the MCT. The degrees of the approximation have also been shown clearly. After simplifying our equation, we can analyze the static result for the glass relaxation. We find that the glass relaxation is ergodic. This is the main result we obtained in this study. In the numerical calculations, we find that the force fluctuation of an individual particle is sensitive to the temperature. The force fluctuation reflects the unbalanced configuration of particles. Thus, The force fluctuation is suggested to be an order parameter for the structural transition in the glass relaxation. This is our second result obtained in this study.\\

Due to the non-linearity, it is difficult to solve the equation numerically. One of the topics in our future study is to find out a proper numerical solution to the equation, by which we can analyze the glass relaxation in a short time range. Besides this topic, extending the equation from the pairwise interaction to the many body interaction is our next goal. Thus, we can complete our theory for the glass relaxation.\\  

The author kindly acknowledges Prof. Ning-Hua Tong from Renmin University of China for discussions.\\    
\appendix
\section{}
In this Appendix, we derive the second term in the right hand side of Eq.(\ref{IIA6}). The method for the derivation can also be applied for the simplification of the second term in the right hand side of Eq.(\ref{IIB9}).\\

Before we start our simplification, we firstly clarify the Theorem of Equipartition Energy(TEE) for easy reading. In the SI units, the TEE reads
$k_BT_{SI}/2=M^{(l)}<[v_x^{(l)'}]^2>/2$ for one degree of freedom along the coordinate axis $x$. The velocity $v_x^{(l)'}$ is for a particle of the $l$-th species. We non-dimensionalize $v_x^{(l)'}$ by the velocity scale $v_{scale}$ to get $v_x^{(l)}$, meaning $v_x^{(l)'}=v_{scale}\cdot v_x^{(l)}$. Substituting the expression of $v_x^{(l)'}$ into the TEE and using the definition of normalized temperature $\eta$, we have $<[v_x^{(l)}]^2>=\eta/\alpha^{(l)}$. The velocity scale  $v_{scale}^2=\epsilon^{(1)}/M^{(1)}$  and the temperature $\eta$ have been defined in Sec.(\ref{IIA}).\\

We correlate the both sides of Eq.(\ref{IIA5}) with the initial density $R_{-\vec{k}}^{(w)}$, and get two terms from the right hand side of Eq.(\ref{IIA5}). We focus on the second term which reads $T_2=\sum_n <(i\vec{k}\cdot \vec{v}_n^{(l)})e^{i\vec{k}\cdot \vec{x}_n^{(l)}}(i\vec{k}\cdot \vec{v}_n^{(l)})R_{-\vec{k}}^{(w)}>/\sqrt{N_l}$. For simplicity, we note $(i\vec{k}\cdot \vec{v}_n^{(l)})(i\vec{k}\cdot \vec{v}_n^{(l)})$ by $L_n$. Thus, the short notation for $T_2$ is $T_2=\sum_n <L_ne^{i\vec{k}\cdot \vec{x}_n^{(l)}}R_{-\vec{k}}^{(w)}>/\sqrt{N_l}$. Note that $L_ne^{i\vec{k}\cdot \vec{x}_n^{(l)}}$ is time dependent while $R_{-\vec{k}}^{(w)}$ is the initial density. For convenience, we note the core $<L_ne^{i\vec{k}\cdot \vec{x}_n^{(l)}}R_{-\vec{k}}^{(w)}>$ of $T_2$ by $T_3$. Our goal is to decouple $<L_n>$ from the remained part of $T_3$, say $<L_ne^{i\vec{k}\cdot \vec{x}_n^{(l)}}R_{-\vec{k}}^{(w)}>\approxeq <L_n><e^{i\vec{k}\cdot \vec{x}_n^{(l)}}R_{-\vec{k}}^{(w)}>$. After the decoupling, the TEE can be applied on $L_n$ directly for simplification to get $<L_n>\approxeq -k^2\eta/\alpha^{(l)}$. Such decoupling can be obtained approximately. In the following, we show the degree of the approximation for the decoupling. \\

We think three variable spaces $P_1$, $P_2$ and $P_3$. The space $P_1$ contains no current $J_{\vec{k}}^{(w)}$ but only products of densities $\rho_{\vec{k}}^{(w)}$ for all the possible slow variables. The spaces $P_2$ and $P_3$ contain products of the currents and the densities. In each product of $P_2$, the number of the currents is odd, while in each product of $P_3$, the number of the currents is even.
\begin{enumerate}
\item{We project the correlation $T_3$ onto the space $P_1$ through the operator $|P_1><P_1|$, leading to $<L_ne^{i\vec{k}\cdot \vec{x}_n^{(l)}}P_1><P_1R_{-\vec{k}}^{(w)}>$. We find that $L_n$ can be decoupled by $<L_ne^{i\vec{k}\cdot \vec{x}_n^{(l)}}P_1>=<L_n><e^{i\vec{k}\cdot \vec{x}_n^{(l)}}P_1>$ because the velocities in $L_n$ are not correlated to $\vec{x}_n^{(l)}$ and densities in $P_1$.}
\item{Similarly, we project $T_3$ onto $P_2$ through the operator $|P_2><P_2|$. In each term of $<L_ne^{i\vec{k}\cdot \vec{x}_n^{(l)}}P_2>$, the number of the velocities is odd. Two velocities are from $L_n$ and the other velocities of odd number are from the odd-number currents $J_{\vec{k}}^{(w)}$ in the $P_2$ space. The ensemble average of velocities of odd number is always zero. Thus, we have $<L_ne^{i\vec{k}\cdot \vec{x}_n^{(l)}}P_2>=0$. We take the form of $<L_ne^{i\vec{k}\cdot \vec{x}_n^{(l)}}P_2>=<L_n><e^{i\vec{k}\cdot \vec{x}_n^{(l)}}P_2>$ for the decoupling. Here, we have utilized $<e^{i\vec{k}\cdot \vec{x}_n^{(l)}}P_2>=0$ due to the odd number of velocities in each product of $P_2$.}
\item{As the same, we project $T_3$ onto the space of $P_3$ through the operator $|P_3><P_3|$. We still treat the factor $<L_ne^{i\vec{k}\cdot \vec{x}_n^{(l)}}P_3>$. The number of the velocities in each term of $P_3$ is even. We take $<v_n^2v_j^2\cdots >$ as an example with $v_n^2$ from $L_n$ and $v_j^2$ from the product of two currents $J_{\vec{k}}^{(w)}$ in $P_3$. $\cdots$ represents the remained part that is not correlated to $v_n^2$. Suppose $n\neq j$ and the $n$-th particle is far from the $j$-th particle, $v_n^2$ is not correlated to $v_j^2$, leading to the decoupling of $<v_n^2v_j^2\cdots>=<v_n^2><v_j^2\cdots>$. Under such condition, $<L_n>$ can be decoupled directly. The strong correlation happens when $n=j$, which brings difficulty for decoupling. In this case, we use the mean thermo speed to replace the velocity in $L_n$ for approximation. With such approximation, we decouple the factor by $<L_ne^{i\vec{k}\cdot \vec{x}_n^{(l)}}P_3>\approxeq -[8k^2\eta/(\pi\alpha^{(l)})][1/3]<e^{i\vec{k}\cdot \vec{x}_n^{(l)}}P_3>$, in which $8\eta/(\pi\alpha^{(l)})$ is for the mean thermo speed and the factor $1/3$ is because the velocities in $L_n$ both are along only one direction $\hat{k}$. By using $<L_n>\approxeq -k^2\eta/\alpha^{(l)}$, we rewrite  $<L_ne^{i\vec{k}\cdot \vec{x}_n^{(l)}}P_3>\approxeq 0.85<L_n><e^{i\vec{k}\cdot \vec{x}_n^{(l)}}P_3>$.}

\item{The three variable spaces $P_1,P_2,P_3$ comprising the full variable space $P$ of the system approximately. The fast variable space constructed by the forces is perpendicular to the velocities and is unnecessary to be considered. Variables faster than the forces are neglected in this approximation. An unit projection operator $|P><P|$ then can be introduced as a function of the operators $|P><P|=a_1|P_1><P_1|+a_2|P_2><P_2|+a_3|P_3><P_3|$. The unit operator inserts into $T_3$ to get 
\begin{align}
&T_3=<L_ne^{i\vec{k}\cdot \vec{x}_n^{(l)}}P><PR_{-\vec{k}}^{(w)}>\nonumber\\
&=<L_n><e^{i\vec{k}\cdot \vec{x}_n^{(l)}}[a_1P_1><P_1+a_2|P_2><P_2|\nonumber\\
&+0.85\times a_3|P_3><P_3]R_{-\vec{k}}^{(w)}>\nonumber\\
&\approxeq a_4<L_n><e^{i\vec{k}\cdot \vec{x}_n^{(l)}}P><PR_{-\vec{k}}^{(w)}>.
\end{align}
In the above treatment, we replace the factor $0.85$ by $1$ because they are very close to each other, and introduce a factor $a_4$ to compensate the deviation due to the replacement. $a_4$ is also close to $1$. 
}
\end{enumerate}
Basing on the above statement, we decouple $<L_n>$ for $T_2$ and have
\begin{align}
&T_2=\frac{\sum_n <L_ne^{i\vec{k}\cdot \vec{x}_n^{(l)}}R_{-\vec{k}}^{(w)}>}{\sqrt{N_l}}\nonumber\\
&\approxeq \frac{a_4\sum_n <L_n><e^{i\vec{k}\cdot \vec{x}_n^{(l)}}|P><P|R_{-\vec{k}}^{(w)}>}{\sqrt{N_l}}\nonumber\\
&=-\frac{a_4k^2\eta}{\alpha^{(l)}}<\rho_{\vec{k}}^{(l)}R_{-\vec{k}}^{(w)}>.
\end{align} 
In the above result, the factor $<L_n>$ has been simplified by the TEE and the definition of density $\rho_{\vec{k}}^{(l)}$ have also been applied. For convenience, the factor $a_4$ is absorbed in the normalized mass $\alpha^{(l)}$ as the effective mass. Finally, we write $T_2=-\frac{k^2\eta}{\alpha^{(l)}}<\rho_{\vec{k}}^{(l)}R_{-\vec{k}}^{(w)}>$ for the simplification.

\section{}
We express $<R_{\vec{k}}^{(l)}R_{-\vec{k}}^{(w)}>$ explicitly as
\begin{align}
<R_{\vec{k}}^{(l)}R_{-\vec{k}}^{(w)}>=\frac{1}{\sqrt{N_lN_w}}<\sum_{i=1}^{N_l}\sum_{j=1}^{N_w}e^{i\vec{k}\cdot (\vec{x}_{i}^{(l)}-\vec{x}_{j}^{(w)})}>.
\end{align}
We extracted the term of $l=w$ and $i=j$ from the above expression to avoid the singularity and denote the remained terms by $S^{(l,w)}$. The term extracted out equals $1$ for $\vec{x}_{i}^{(l)}=\vec{x}_{j}^{(w)}$. The expression then is rewritten as $<R_{\vec{k}}^{(l)}R_{-\vec{k}}^{(w)}>=\delta_{l,w}+S^{(l,w)}_{i,j}$. Here, $\delta_{l,w}$ is the Kronecker delta function, which equals $1$ if $l=w$ and zero if $l\neq w$. In $S^{(l,w)}$, it is restricted by $\vec{x}_{i}^{(l)}\neq\vec{x}_{j}^{(w)}$. Then We go further to get
\begin{align}
\label{B2}
&S^{(l,w)}=\frac{1}{\sqrt{N_lN_w}}<\sum_{n=1}^{N_l}\sum_{m=1}^{N_w}e^{i\vec{k}\cdot (\vec{x}_{n}^{(l)}-\vec{x}_{m}^{(w)})}>\nonumber\\
&=\frac{1}{\sqrt{N_lN_w}}<\sum_{n=1}^{N_l}\sum_{m=1}^{N_w}\int d \vec{r}e^{-i\vec{k}\cdot \vec{r}}\delta(\vec{r}+\vec{x}_{n}^{(l)}-\vec{x}_{m}^{(w)})>\nonumber\\
&=\sqrt{\frac{N_l}{N_w}}\int d \vec{r}e^{-i\vec{k}\cdot \vec{r}}\rho_wg_{_{lw}}(r)
\end{align}
with $g_{_{lw}}(r)=<\sum_{n=1}^{N_l}\sum_{m=1}^{N_w}\delta(\vec{r}+\vec{x}_{n}^{(l)}-\vec{x}_{m}^{(w)})>/(N_l\rho^{(w)})$ the radial distribution function. The factor $\sqrt{N_l/N_w}$ is replaced by the ratio of the particle density $\sqrt{\rho^{(l)}/\rho^{(w)}}$, and then we reformulate $S^{(l,w)}$ to be
\begin{align}
&S^{(l,w)}=\sqrt{\rho^{(l)}\rho^{(w)}}\int d \vec{r}e^{-i\vec{k}\cdot \vec{r}}g_{_{lw}}(r).
\end{align} 
Since the system is homogeneous, the integration $\int d \vec{r}e^{-i\vec{k}\cdot \vec{r}}$ is replaced by $(4\pi/k)\int d r~ r\sin(kr)$. After the integration, the singularity $\delta(k)$ should be subtracted. We need replace $g_{_{lw}}(r)$ by $g_{_{lw}}(r)-1$. Finally, we have Eq.(\ref{IID11}).\\

The explicit form for $<\mathcal{T}_{\vec{k}}^{(l)}\mathcal{T}_{-\vec{k}}^{(w)}>$ is
\begin{align}
\label{TT}
&<\mathcal{T}_{\vec{k}}^{(l)}\mathcal{T}_{-\vec{k}}^{(w)}>=<\sum_{p}(i\hat{k}\cdot T_{\vec{k}}^{(l,p)})\sum_{q}(-i\hat{k}\cdot T_{-\vec{k}}^{(w,q)})>\nonumber\\
&=\frac{1}{\sqrt{N_l N_w}}<\sum_{n,m} e^{i\vec{k}\cdot (\vec{x}_n^{(l)}-\vec{x}_m^{(w)})}(\hat{k}\cdot\ddot{\vec{x}}^{(l)}_n) (\hat{k}\cdot\ddot{\vec{x}}^{(w)}_m)>.
\end{align}
Here, $\ddot{\vec{x}}^{(l)}_n$ is the acceleration for the $n$-th particle of the $l$-th species, which equals
\begin{align}
\label{B5}
\ddot{\vec{x}}^{(l)}_n=-\frac{1}{\alpha^{(l)}}\sum_{p,s}\frac{\partial V(\vec{x}_n^{(l)},\vec{x}_s^{(p)})}{\partial \vec{x}_n^{(l)}}.
\end{align} 
We split the expression of $<\mathcal{T}_{\vec{k}}^{(l)}\mathcal{T}_{-\vec{k}}^{(w)}>$ into two terms. The first term $T_1$ is for the case of $l=w$ and $m=n$, which reads
\begin{align}
T_1=\frac{1}{N_l}<\sum_{n} (\hat{k}\cdot\ddot{\vec{x}}^{(l)}_n)^2>\delta_{l,w}.
\end{align}
The remained parts are for the second term $T_2$, which is restricted by $\vec{x}_n^{(l)}\neq\vec{x}_m^{(w)}$. Similar to Eq.(\ref{B2}), $T_2$ is expressed as
\begin{align}
&T_2=\frac{1}{\sqrt{N_l N_w}}<\sum_{n,m} e^{i\vec{k}\cdot (\vec{x}_n^{(l)}-\vec{x}_m^{(w)})}(\hat{k}\cdot\ddot{\vec{x}}^{(l)}_n) (\hat{k}\cdot\ddot{\vec{x}}^{(w)}_m)>\nonumber\\
&=\frac{4\pi \sqrt{\rho^{(l)}\rho^{(w)}}}{k}\int dr~r \sin(kr)h_{lw}(r)
\end{align}
with $h_{lw}(r)=<\sum_{n,m}\delta(\vec{r}+\vec{x}_n^{(l)}-\vec{x}_m^{(w)})(\hat{k}\cdot\ddot{\vec{x}}^{(l)}_n) (\hat{k}\cdot\ddot{\vec{x}}^{(w)}_m)>/(N_l\rho^{(w)})$. The MD simulations for $g_{_{lw}}(r)$ and $h_{_{lw}}(r)$ have been specified in Sec.(\ref{IID}).

\section{}
We have the definition
\begin{align}
\label{C1}
F_{\vec{k}}^{(l,p)}=-\frac{1}{\alpha^{(l)}\sqrt{N_l}}\sum_{n,m} e^{i\vec{k}\cdot \vec{x}_n^{(l)}}[\frac{\partial V(\vec{x}_n^{(l)},\vec{x}_m^{(p)})}{\partial \vec{x}_n^{(l)}}]
\end{align}
in Eq.(\ref{IIA4}). The particles are not coincident in the MD simulations due to the steric effect. Therefore, $\vec{x}_m^{(p)} \neq \vec{x}_n^{(l)}$ will be held automatically in MD simulations. The potential $V(\vec{x}_n^{(l)},\vec{x}_m^{(p)})$ is functional of $\vec{x}_n^{(l)}-\vec{x}_m^{(p)}$. We make a Fourier transformation on the potential to get
\begin{align}
\label{C2}
V(\vec{x}_n^{(l)},\vec{x}_m^{(p)})=\frac{1}{V_{\vec{k}}}\int e^{-i \vec{k}_1\cdot (\vec{x}_n^{(l)}-\vec{x}_m^{(p)})} B_{\vec{k}_1} d\vec{k}_1
\end{align}
with $B_{\vec{k}_1}$ the Fourier component at $\vec{k}_1$ and $V_{\vec{k}}$ the volume for the integration in the reciprocal space. By substituting Eq.(\ref{C2}) into Eq.(\ref{C1}), we rewrite $F_{\vec{k}}^{(l,p)}$ in the form of
\begin{align}
F_{\vec{k}}^{(l,p)}=\frac{1}{V_{\vec{k}}}\int \left[\sum_{n}e^{i (\vec{k}-\vec{k}_1)\cdot \vec{x}_n^{(l)}}\right]C_{\vec{k}_1} \left[\sum_{m} e^{i\vec{k}_1\cdot \vec{x}_m^{(p)}}\right] d\vec{k}_1
\end{align}
by absorbing the coefficients in  $C_{\vec{k}_1}$. Note that $\rho_{\vec{k}-\vec{k}_1}^{(l)}=\sum_{n}e^{i (\vec{k}-\vec{k}_1)\cdot \vec{x}_n^{(l)}}/\sqrt{N_l}$, and $\rho_{\vec{k}_1}^{(p)}=\sum_{m} e^{i\vec{k}_1\cdot \vec{x}_m^{(p)}}/\sqrt{N_p}$. Then, we obtain
\begin{align}
\mathcal{F}_{\vec{k}}^{(l)}=\sum_{p}(i\hat{k}\cdot F_{\vec{k}}^{(l,p)})=\sum_{p}\frac{1}{V_{\vec{k}}}\int  B_{\vec{k},\vec{k}_1}^{(l,p)} \rho_{\vec{k}-\vec{k}_1}^{(l)}\rho_{\vec{k}_1}^{(p)}d \vec{k}_1.
\end{align}
with $B_{\vec{k},\vec{k}_1}^{(l,p)}$ for the coefficients. We can not derive an analytical expression for $B_{\vec{k},\vec{k}_1}^{(l,p)}$, but can use MD simulations to access it, which has been shown in Sec.(\ref{IIF}).\\

\section{}
We have used $\ddot{\vec{x}}^{(l)}_n$ to represent the acceleration for the $n$-th particle of the $l$-th species. We define further a quantity $\ddot{\vec{x}}^{(l,p)}_{n,m}$ as the component of $\ddot{\vec{x}}^{(l)}_n$ acted by the $m$-th particle of the $p$-th species only, taking the form of $\ddot{\vec{x}}^{(l,p)}_{n,m}=-[\frac{\partial V(\vec{x}_n^{(l)},\vec{x}_m^{(p)})}{\partial \vec{x}_n^{(l)}}]/\alpha^{(l)}$. Then we have $\ddot{\vec{x}}^{(l)}_n=\sum_{p,m}\ddot{\vec{x}}^{(l,p)}_{n,m}$. We define $\ddot{\vec{x}}^{(l,l)}_{n,n}=0$ due to no physical effect. We introduce a pair of Fourier transformation
\begin{align}
\label{D1}
&\sum_m \ddot{\vec{x}}^{(l,p)}_{n,m} e^{-i\vec{k}_1\cdot \vec{x}_m^{(p)}}=y^{(l,p)}_{n,\vec{k}_1},\\
\label{D2}
&\ddot{\vec{x}}^{(l,p)}_{n,m} =\frac{1}{V_{\vec{k}}}\int y^{(l,p)}_{n,\vec{k}_1}e^{i\vec{k}_1\cdot \vec{x}_m^{(p)}}d\vec{k}_1.
\end{align}
For the above Fourier transformation, we have used the result that $\int e^{i\vec{k}_1\cdot(\vec{x}_n^{(p)}-\vec{x}_m^{(p)})}d \vec{k}_1$ equals $0$ for $m\neq n$ and equals $V_{\vec{k}}$ for $m=n$. Then we have
\begin{align}
&\mathcal{F}_{\vec{k}}^{(l)}=\sum_{p}(i\hat{k}\cdot F_{\vec{k}}^{(l,p)})=\frac{i}{ \sqrt{N_l}}\sum_{n} e^{i\vec{k}\cdot \vec{x}_n^{(l)}}(\hat{k}\cdot\ddot{\vec{x}}^{(l)}_{n})\nonumber\\
&=\frac{i}{ \sqrt{N_l}}\sum_{n}\left\{\sum_{p,m} e^{i\vec{k}\cdot \vec{x}_n^{(l)}}(\hat{k}\cdot\ddot{\vec{x}}^{(l,p)}_{n,m})\right\}\nonumber\\
&=\sum_{p}\frac{1}{V_{\vec{k}}}\int \left\{\frac{i}{ \sqrt{N_l}} \sum_{n,m} e^{i\vec{k}\cdot \vec{x}_n^{(l)}}(\hat{k}\cdot  y^{(l,p)}_{n,\vec{k}_1})e^{i\vec{k}_1\cdot \vec{x}_m^{(p)}}\right\}d\vec{k}_1
\end{align}
by using Eq.(\ref{IIA4}), Eq.(\ref{B5}), Eq.(\ref{D1}) and Eq.(\ref{D2}). Comparing to  $\mathcal{F}_{\vec{k}}^{(l)}=\sum_{p}\frac{1}{V_{\vec{k}}}\int \left\{ B_{\vec{k},\vec{k}_1}^{(l,p)} \rho_{\vec{k}-\vec{k}_1}^{(l)}\rho_{\vec{k}_1}^{(p)}\right\}d \vec{k}_1$ in Eq.(\ref{IIE13}) component by component, we obtain Eq.(\ref{IIF20}) as
\begin{align}
\label{D4}
\mathcal{I}^{(l,p)}_{\vec{k}, \vec{k}_1}&=\frac{i}{\sqrt{N_l}} \sum_{n,m} e^{i\vec{k}\cdot \vec{x}_n^{(l)}}(\hat{k}\cdot  y^{(l,p)}_{n,\vec{k}_1})e^{i\vec{k}_1\cdot \vec{x}_m^{(p)}}\nonumber\\
&= B_{\vec{k},\vec{k}_1}^{(l,p)} \rho_{\vec{k}-\vec{k}_1}^{(l)}\rho_{\vec{k}_1}^{(p)}
\end{align}
We substitute Eq.(\ref{D1}) into the first line of Eq.(\ref{D4}) and express $y^{(l,p)}_{n,\vec{k}_1}$ in the term of $ \ddot{\vec{x}}^{(l,p)}_{n,m}$ to make $\mathcal{I}^{(l,p)}_{\vec{k}, \vec{k}_1}$ accessible by MD simulations. The quantity $\mathcal{J}_1$ then is
\begin{align}
&\mathcal{J}_1=<\mathcal{I}^{(l,p)}_{\vec{k}, \vec{k}_1}\mathcal{I}^{(w,q)}_{-\vec{k}, -\vec{k}_1}>\nonumber\\
&=\sqrt{\frac{N_pN_q}{N_lN_w}}\sum_{n,u}^{s,v}<(\hat{k}\cdot  \ddot{\vec{x}}^{(l,p)}_{n,u} )(\hat{k}\cdot \ddot{\vec{x}}^{(w,q)}_{s,v} )\nonumber\\
&\times e^{i\vec{k}\cdot (\vec{x}_n^{(l)}-\vec{x}_s^{(w)})}e^{i\vec{k}_1\cdot (\vec{x}_v^{(q)}-\vec{x}_u^{(p)})}R_{\vec{k}_1}^{(p)}R_{-\vec{k}_1}^{(q)}>.
\end{align}
If the two acceleration components $\ddot{\vec{x}}^{(l,p)}_{n,u}$ and $\ddot{\vec{x}}^{(w,q)}_{s,v}$ are independent to each other, the ensemble average of the two acceleration components equals zero. Thus, we consider only two cases in which $\ddot{\vec{x}}^{(l,p)}_{n,u}$ and $\ddot{\vec{x}}^{(w,q)}_{s,v}$ are correlated for nonzero of $\mathcal{J}_1$. In the first case, $\vec{x}_v^{(q)}=\vec{x}_u^{(p)}$ and $\vec{x}_n^{(l)}=\vec{x}_s^{(w)}$ are held. In this case, we have
\begin{align}
&\mathcal{J}_1=<\mathcal{I}^{(l,p)}_{\vec{k}, \vec{k}_1}\mathcal{I}^{(w,q)}_{-\vec{k}, -\vec{k}_1}>\nonumber\\
&=\frac{N_p}{N_l}\left<\sum_{n,u}(\hat{k}\cdot  \ddot{\vec{x}}^{(l,p)}_{n,u} )^2R_{\vec{k}_1}^{(p)}R_{-\vec{k}_1}^{(p)}\right>\delta_{p,q}\delta_{l,w}.
\end{align}
The last case is for $l\neq w$, in which $\vec{x}_s^{(w)}=\vec{x}_u^{(p)}$ and $\vec{x}_n^{(l)}=\vec{x}_v^{(q)}$ are held for the nonzero $\mathcal{J}_1$. In the last case, we have
\begin{align}
&\mathcal{J}_1=<\mathcal{I}^{(l,p)}_{\vec{k}, \vec{k}_1}\mathcal{I}^{(w,q)}_{-\vec{k}, -\vec{k}_1}>\nonumber\\
&=-\left<\sum_{n,s}(\hat{k}\cdot  \ddot{\vec{x}}^{(l,w)}_{n,s} )^2e^{i(\vec{k}+\vec{k}_1)\cdot (\vec{x}_n^{(l)}-\vec{x}_s^{(w)})}R_{\vec{k}_1}^{(w)}R_{-\vec{k}_1}^{(l)}\right>\delta_{p,w}\delta_{l,q}
\end{align}
for $l\neq w$. In this study, we focus on the first case only.\\

\section{}
According to the definition $\mathcal{T}_{\vec{k}}^{(l)}=\frac{1}{\sqrt{N_l}}\sum_{n} e^{i\vec{k}\cdot \vec{x}_n^{(l)}}(i\hat{k}\cdot \ddot{\vec{x}}^{(l)}_n)$, we have
\begin{align}
\label{E1}
&\sqrt{N_lN_w}<\mathcal{T}_{\vec{k}}^{(l)}R_{-\vec{k}}^{(w)}>\nonumber\\
&=<i\hat{k}\cdot(\sum_n e^{i\vec{k}\cdot \vec{x}_n^{(l)}}\ddot{\vec{x}}_n^{(l)})\sum_m e^{-i\vec{k}\cdot \vec{x}_m^{(w)}}>\nonumber\\
&=-i\hat{k}\cdot<\sum_n \dot{\vec{x}}_n^{(l)}\frac{d}{dt}\left(e^{i\vec{k}\cdot \vec{x}_n^{(l)}}\sum_m  e^{-i\vec{k}\cdot \vec{x}_m^{(w)}}\right)>\nonumber\\
&=k\frac{\eta}{\alpha^{(l)}}\left\{\sqrt{N_lN_w}<R_{\vec{k}}^{(l)}R_{-\vec{k}}^{(w)}>-\delta_{w,l}N_l\right\}.
\end{align}
From the second line to the third line, we have used the formula $<d(\dot{x}y)/dt>=0=<\ddot{x}y>+<\dot{x}\dot{y}>$. Finally, we apply the TEE to get the fourth line.\\

According to Eq.(\ref{IIA6}), the initial value is given as
\begin{align}
\ddot{\mathcal{G}}_{\vec{k},0}^{(l,w)}=\frac{<\mathcal{T}_{\vec{k}}^{(l)}R_{-\vec{k}}^{(w)}>}{\eta k<R^{(l)}_{\vec{k}}R^{(w)}_{-\vec{k}}>}-\frac{1}{ \alpha^{(l)}}\mathcal{G}_{\vec{k},0}.
\end{align}
By using Eq.(\ref{E1}), we have the initial condition 
\begin{align}
\ddot{\mathcal{G}}_{\vec{k},0}^{(l,w)}=-\frac{1}{ \alpha^{(l)}}\frac{\delta_{w,l}}{<R_{\vec{k}}^{(l)}R_{-\vec{k}}^{(w)}>}.
\end{align}
Note that in the definition of $\mathcal{G}_{\vec{k}}^{(l,w)}$ in Eq.(\ref{IIC10}), a factor $\sqrt{\eta}k$ has been absorbed in time $t$.


\end{document}